\documentclass[journal]{IEEEtran}
\usepackage{cite}
\usepackage{color}
\usepackage{amsmath,amssymb}

\DeclareMathOperator*{\argmax}{arg\,max}
\newcommand{\cev}[1]{\reflectbox{\ensuremath{\vec{\reflectbox{\ensuremath{#1}}}}}}

\usepackage{algorithm}
\usepackage{algpseudocode}
\makeatletter
\def\BState{\State\hskip-\ALG@thistlm}
\makeatother

\algdef{SE}[SUBALG]{Indent}{EndIndent}{}{\algorithmicend\ }%
\algtext*{Indent}
\algtext*{EndIndent}
\newcounter{rtaskno}

\usepackage{hyperref}
\usepackage{amssymb}

\usepackage{graphicx,caption}

\usepackage{array}
\usepackage{lipsum}
\usepackage{float}
\usepackage{rotating}

\newcommand{\hp}{\overset{\rightharpoonup}}

\allowdisplaybreaks

\begin{document}
%

\thispagestyle{empty} 
\noindent
\begin{center}
    \large\textbf{Copyright Notice}
\end{center}

\vspace{2cm}

\noindent
\copyright\ 2018 IEEE. Personal use of this material is permitted. Permission from IEEE must be obtained for all other uses, in any current or future media, including reprinting/republishing this material for advertising or promotional purposes, creating new collective works, for resale or redistribution to servers or lists, or reuse of any copyrighted component of this work in other works.

\vspace{1cm}

\noindent
\textbf{To cite this article:} \\
S. Lin, ``Reverberation-robust localization of speakers using distinct speech onsets and multichannel cross correlations,'' in \textit{IEEE/ACM Transactions on Audio, Speech, and Language Processing}, vol. 26, no. 11, pp. 2098--2111, 2018.

\vspace{1cm}

\noindent
\textbf{Official Version of Record:} \\
The final version of this article is available at: 

\url{https://ieeexplore.ieee.org/abstract/document/8409999/}

\url{https://doi.org/10.1109/TASLP.2018.2854871}

\vspace{2cm}
\noindent
\textbf{Dr. Shoufeng Lin has been a Senior Member of IEEE since 2020.}

\newpage 
\setcounter{page}{1} 

\title{Reverberation-Robust Localization of Speakers Using Distinct Speech Onsets and Multi-channel Cross-Correlations}

\author{Shoufeng~Lin,~\IEEEmembership{Member,~IEEE}
\thanks{Shoufeng Lin is {\color{black}currently}
with the {\color{black}School of Electrical Engineering, Computing and Mathematical Sciences}, Curtin University, Bentley, Western Australia. E-mail: shoufeng.lin@postgrad.curtin.edu.au; ee.linsf@gmail.com.}
}

\maketitle

\begin{abstract}
Many speaker localization methods can be found in the literature. However, speaker localization under strong reverberation still remains a major challenge in the real-world applications. 
This paper proposes {\color{black}two} algorithm{\color{black}s} for localizing speakers using microphone array recordings of reverberated sounds. 
To separate concurrent speakers, {\color{black}the first algorithm} decompose{\color{black}s} microphone signals spectrotemporally into subbands via an auditory filterbank. 
To suppress reverberation, we propose a novel speech onset detection approach derived from the speech signal and impulse response models, and further propose to formulate the multi-channel cross-correlation coefficient (MCCC) of encoded speech onsets in each subband. The subband results are combined to estimate the directions-of-arrival (DOAs) of speakers. 
The {\color{black}second algorithm} extends the generalized cross-correlation - phase transform (GCC-PHAT) method by using redundant information of multiple microphones to address the reverberation problem. 
The proposed methods have been evaluated under adverse conditions using not only simulated signals (reverberation time $T_{60}$ of up to $1$s) but also recordings in a real reverberant room ($T_{60} \approx 0.65$s). Comparing with some state-of-the-art localization methods, experimental results confirm that the proposed methods can reliably locate static and moving speakers, in presence of reverberation. 

\end{abstract}


%
\IEEEpeerreviewmaketitle

\section{Introduction}
\label{intro}
%

\IEEEPARstart{L}{ocalizing} speakers using signals recorded by microphones is an important problem in audio signal processing, especially for a series of applications such as speech acquisition, separation, recognition and transcription, as well as speaker tracking, automatic camera steering for online lectures and video-conferencing in smart environments. 
In this paper, we focus the study on the problem of localizing speakers in reverberant conditions. 


Numerous studies can be found on sound source localization, but most still have problems when dealing with concurrent speakers and strong reverberation{\color{black}, especially when speakers are moving}. 
In general, speaker localization methods can be classified into three categories, namely the subspace-based techniques, the steered-response beamformers, and the approaches based on time-difference-of-arrival (TDOA) estimation.  
The subspace-based methods, including the popular MUSIC (multiple signal classification) \cite{schmidt1986multiple} and ESPRIT (estimation of signal parameters via rotational invariance techniques) \cite{roy1989esprit} and their variants \cite{rao1989performance, pesavento2000unitary}, can provide {\color{black}good} direction-of-arrival (DOA) resolutions for spatially uncorrelated narrowband sources. Although their wideband extensions are available \cite{ottersten1990direction, teutsch2006acoustic}, most of them still have the difficulty in handling spatially coherent sources, i.e. strong sound reflections. 
The steered-response power (SRP) beamforming methods are an intuitive solution for DOA estimation, by steering the beam over DOAs and analyz{\color{black}ing} the corresponding response power. Besides the many developed wideband acoustic beamformers, recently the circular harmonics beamformer (CHB) \cite{parthy2011comparison, tiana2010beamforming, torres2012robust} is getting attention in the wideband source localization applications, and in particular, a time-frequency (TF) - CHB \cite{torres2012robust} was reported to have outperformed the eigenbeam (EB) - ESPRIT \cite{teutsch2006acoustic} and the delay-and-sum beamformer (DSB) in localizing multiple sources under high reverberation and noise. 
Although quite straightforward, due to its beamwidth, a wideband beamformer usually has comparatively limited DOA resolution versus the number of sensors, and locations of the peak steered-response power may be easily offset by the beam pattern and concurrent sources.

Most existing TDOA estimation algorithms are built upon the generalized cross-correlation (GCC) \cite{knapp1976generalized} or the eigenvalue decomposition (EVD) \cite{benesty2000adaptive} method. 
The GCC method, especially with the classical phase transform (PHAT) processor, has been very popular but is often regarded sensitive to reverberation due to the free-field plane-wave model it assumes. 
Thus the multi-channel cross-correlation coefficient (MCCC) method has been developed \cite{benesty2004time}, by making use of redundant information from multiple microphones. However, the microphone signals still contain reflection components, and MCCC uses cross-correlation coefficients of all microphone pairs including those far apart, which can be degraded by spatial alias. 
The EVD method however, may not directly handle the case of multiple concurrent speakers, and can have problems when impulse responses of speaker-microphone channels have common zeros \cite{benesty2000adaptive}. 
%
{\color{black} To address the reverberation problem, many works exploit the ``precedence effect'' by extracting speech onsets\footnote{\color{black}Here the speech onset refers to a sudden increase in the speech envelope, while the speech offset denotes a decay of the speech energy \cite{dixon2006onset, tho2014robust}.} 
as the reliable localization cues \cite{rakerd1986localization, huang1997sound, dixon2006onset, smith2007determining, kuhne2009robust, lee2010multiple, tho2014robust}. 
On the one hand, some onset detection methods are formulated in the short time Fourier transform (STFT) domain \cite{dixon2006onset, kuhne2009robust, tho2014robust}, by calculating the phase deviation or magnitude rise. 
On the other hand, filterbank-based methods can also be found in the literature \cite{huang1997sound, smith2007determining, lee2010multiple, plinge2010robust}.
An ``echo-free onset detection model'' was developed in \cite{huang1997sound} and further implemented in \cite{lee2010multiple}. 
However, most parameters of these onset detection methods were empirical.}
Inspired by the computational auditory scene analysis (CASA) techniques \cite{ASAbregman, CASAwang}, 
{\color{black}a two-microphone localization algorithm was developed, which encodes the bandpass-filtered signal to imitate the ``auditory nerve spikes'' and compares onset spike times to estimate time differences\cite{smith2007determining}.} 
A ``Neuro-Fuzzy'' multi-speaker localization method was recently developed \cite{plinge2010robust}, which combines cross-correlation coefficients of {\color{black}pulse-}encoded peaks over moving averages of subband signals and shows robustness against reverberation. 
However, {\color{black}both \cite{smith2007determining} and \cite{plinge2010robust} rely} on psychoacoustic experimental inferences {\color{black}(}e.g. the ``glimpses'' \cite{cooke2006glimpsing}, ``precedence effect'' \cite{litovsky1999precedence} and neural spikes generation \cite{meddis1986simulation, CASAwang}{\color{black})}, rather than signal models and mathematical derivations. 

In this paper, CASA-inspired but mathematically motivated, we propose a novel speaker localization method using multi-channel cross-correlation coefficients of distinct speech onsets. 
We first model the speech signals and acoustic room impulse responses (RIR), and decompose the speech mixtures into subbands based on the sparsity assumption of speech signals on TF domain \cite{yilmaz2004blind}, so that speech signal components from separate speakers do not significantly overlap in each subband. 
Then based on the signal and room models, we propose and derive a novel approach to detect and encode the distinct speech onsets, which provides reliable direct-path components while suppressing random reflections due to reverberation. 
We further formulate MCCC of the encoded speech onsets with spatial alias avoided and combine subband results for estimating DOAs of speakers. 
This proposed localization method is referred to as the Onset-MCCC method in this paper. 
Moreover, {\color{black}we also provide discussions of the signal models and algorithms in the short-time Fourier transform (STFT) domain. There} we extend the GCC-PHAT method using multiple closely placed microphones for speaker localization in presence of reverberation, motivated by the idea of using redundant information of multiple microphones to suppress reflections. We refer to this method as the multi-channel cross-correlation (MCC)-PHAT method in this paper. 
The performance of the proposed Onset-MCCC and MCC-PHAT methods are compared with the Neuro-Fuzzy, TF-CHB and EB-ESPRIT methods. 
%
%


The remainder of this paper is organized as follows. 
Section \ref{section:decomp} provides the speech signal and RIR models and the decomposition of speech mixtures, 
Section \ref{section:onset} {\color{black}proposes the} speech onset detection and encoding methods, and Section \ref{section:localization} {\color{black}proposes the} Onset-MCCC method{\color{black}. Section \ref{sec:STFT} presents the discussions in STFT domain and the MCC-PHAT method.}
Performance evaluations are demonstrated in Section \ref{section:performance}. 
Conclusions are given in Section \ref{section:conclusion}.


\section{\color{black} Signal Models and Subband Decomposition}
\label{section:decomp}

%
%

\subsection{Speech Signal Model} \label{section:speechmodel}
Most speech signals are composed of the voiced, unvoiced sounds and silences, where the voiced sounds dominate the signal energy in general. Based on the source excitation - vocal tract models for the process of speech production \cite{deller1993discrete}, as well as the amplitude-modulation (AM) and frequency modulation (FM) structure \cite{maragos1993energy}, for each short segment of voiced sounds, we can use the harmonic model:
\begin{equation}
\label{eq:speechModel}
s_q(t) = \sum \limits _{{\hbar}=1} ^{H_q} s^{(\hbar)}_q(t) , 
\end{equation}
\begin{equation}\label{eq:speechModel2}
s^{(\hbar)}_q(t) = A^{(\hbar)}_{q}(t) \cdot \cos \big( {\hbar} \cdot \omega_q \cdot t + \phi^{({\hbar})}_{q}(t) \big) ,
\end{equation}
where $t\in \mathbb{R}$ is the continuous time, $s_q(t)$ is the speech signal from the $q$-th speaker, $s^{(\hbar)}_q(t)$ the $\hbar$-th harmonic of speaker $q$, $q = 1,...,Q $, integer $Q\geq 1$ the number of concurrent speakers, integer ${\hbar}$ the order of harmonics for a speaker, integer $H_q$ the maximum order of harmonics for speaker $q$, $A^{({\hbar})}_{q}(t) \geq 0$ the envelope of each harmonic, $\phi^{({\hbar})}_{q}(t) \in \mathbb{R}$ the phase {\color{black}(which is assumed constant for a short interval of time)}, $\omega_q > 0$ the {\color{black}angular} fundamental frequency{\color{black}, which is usually different for concurrent speakers}. Compared to the modulating harmonic frequency, {\color{black}the bandwidth of $A^{({\hbar})}_{q}(t)$ is usually small. For the speech onset, $A^{({\hbar})}_{q}(t)$ is a rising ramp.}
%

\subsection{Mixture of Concurrent Speakers}
In reverberant environments, e.g. an enclosed room, signals acquired by the $i$-th microphone are mixtures of reverberated speech signals from speakers and noise: 
\begin{equation} \label{eq:micSig}
{x}_i(t) 
= \sum \limits _{q=1} ^{Q} s_q(t) \ast \hat{\mathrm{h}}_{q i}(t)  + n_i(t)
,
\end{equation}
where the convolution operation is denoted as $\ast$, the additive noise at the $i$-th microphone as $n_i(t)$, $i=1,...,I$, and integer $I \geq 2$ is the number of microphones. $\hat{\mathrm{h}}_{qi}(t) \in \mathbb{R}$ is the acoustic RIR from the $q$-th speaker to the $i$-th microphone.

\subsection{Acoustic Reverberation}
\label{section:reverb}

{\color{black}Acoustic RIR is often modeled in three parts, viz. the direct-path, early reflections and the late (diffuse) reverberation \cite{begault20003, lehmann2010diffuse}.
The early part of RIR typically contains the direct-path and some discrete reflections, while the diffuse part is normally distributed with exponentially decaying envelope. This paper assumes that sound sources are not located too close to reflective surfaces, thus the early reflections are negligible \cite{lollmann2010improved}:} 
%
\begin{equation}
\label{eq:reverbIR}
{\mathrm{h}_{qi}}(t) = 
\begin{cases}
{\mathrm{h}_{d_{qi}}} , ~ t = t_{d_{qi}} \\
{\mathrm{h}_{d_{qi}}} \cdot \nu_{qi}(t -t_{d_{qi}}) \cdot 10^{-{3} \cdot \frac{ t -t_{d_{qi}} }{T_{60} }}, \; ~ t \geq t_{d_{qi}} + \tau_{{qi}} \\
0, ~\mathrm{otherwise} 
\end{cases}
\end{equation}
where $t_{d_{qi}} \in \mathbb{R}$ is the sound travelling time via the direct-path, ${\mathrm{h}_{d_{qi}}} ~({\mathrm{h}_{d_{qi}}} > 0)$ is the magnitude of the direct-path impulse response, $\nu_{qi}(\cdot) \sim \mathcal{N}(0,1)$ are random variables that follow the normal distribution with zero mean and variance of 1, and $\tau_{{qi}} $ ($\tau_{{qi}}  > 0 $) is the time delay for the first (usually strongest) {\color{black}diffuse} reflection to arrive after the direct-path. 
In practice, the exact value of $T_{60}$\footnote{Reverberation of an enclosed environment is usually characterized with $T_{60}$, which is the time required for sound to decay by $60$dB.} of the environment is often unknown \textit{a priori}, but it can be measured \cite{schroeder1965response} or estimated from sound recordings \cite{lollmann2010improved}.

Using the mathematical expectation $\mathbb{E}(\cdot)$, it is easy to check that $\mathbb{E}(\mathrm{h}_{{qi}}^2(t_{d_{qi}}+T_{60})) = \mathbb{E}(\mathrm{h}_{{qi}}^2(t_{d_{qi}}))\cdot 10^{-6}$, which is $60$dB below the direct-path response $\mathbb{E}(\mathrm{h}_{{qi}}^2(t_{d_{qi}})) = \mathrm{h}_{d_{qi}}^2$.

Note that $\tau_{{qi}}$ in (\ref{eq:reverbIR}) is usually over a few milliseconds {\color{black}(ms)} so that one can {\color{black}make a distinction between} the direct-path and reflections. 
This can be connected with some psychoacoustic observations that {\color{black}when there is no time delay, a human listener may fuse two click sounds into a perceived ``phantom'' source between them, which shifts towards the leading sound as time delays increase to 1ms. With delays between 1ms and the ``echo threshold'' (a few ms), one may hear only the leading sound \cite{litovsky1999precedence, begault20003}.} 
{\color{black}Therefore, the RIR model assumes that the early reflections within a few ms following the direct-path are negligible.}

\subsection{Subband Decomposition {\color{black}via Auditory Filterbank}}
\label{sec:subbandDecomp}
Based on the TF sparsity assumption \cite{yilmaz2004blind} {\color{black}and the harmonic structure of speech signal (\ref{eq:speechModel})}, to separate signal components from different speakers, signals of each microphone can be decomposed spectrotemporally via an auditory filterbank so that speech components from separate speakers do not overlap much in each subband {\color{black}(see e.g. \cite{patterson1987efficient, holdsworth1988implementing, smith2007determining, CASAwang, plinge2010robust})}
\begin{equation}
\label{eq:micSignal}
{x}_i^{(b)}(t) = x_i(t) \ast g^{(b)}(t) ,
\end{equation}
where ${x}_i^{(b)}(t)$ denotes the decomposed signals from the $i$-th microphone in subband $b$, $b=1,\dots, N_b$, integer $N_b$ is the total number of subbands, and 
$g^{(b)}(t)$ is the filter impulse response of subband $b$, which is aligned in time between subbands. Common auditory filters include the gammatone filter \cite{patterson1987efficient,holdsworth1988implementing, CASAwang}, gammachirp filter, etc. 

From (\ref{eq:micSig}) {\color{black}and}  (\ref{eq:micSignal}), {\color{black}for the duration} when the noise is small in the particular subband {\color{black}(the SNR estimation is outside of the scope of this paper)}, the decomposed signals in subband $b$ become:
\begin{equation}
\begin{aligned} \label{eq:fbsig}
{x}_{i}^{(b)}(t)  
%
& \approx  \sum \limits _{q=1} ^{Q} s_q(t) \ast \mathrm{h}_{q i}(t)  \ast g^{(b)}(t) . 
%
\end{aligned}
\end{equation}
{\color{black}Moreover, for the harmonic component $\hbar$ of the $q$-th speaker that falls within the passband of subband $b$, (\ref{eq:fbsig}) further simplifies to} 
\begin{equation} \label{eq:fbsig1}
{x}_{i}^{(b)}(t) \approx s^{(\hbar)}_q(t) \ast \mathrm{h}_{qi}(t) \ast g^{(b)}(t) , 
\end{equation}
using the commutativity and associativity properties of convolution, and the frequency selectivity of the filterbank. 

\section{Detecting and Encoding Speech Onsets}
\label{section:onset}
In reverberant environments, the subband signals ${x}_{i}^{(b)}(t) $ in (\ref{eq:fbsig}) are mixtures of direct-path and reflection components. 
We can thus find locations of speakers via detecting the direct-paths and suppressing random reflections.
In this section, we propose a novel approach based on the models of speech mixtures and acoustic reverberation discussed in Section \ref{section:decomp}.

\subsection{Speech Onsets, Direct-paths and Reflections}
\label{section:onsets_dir}

Suppose there is an arbitrary distinct speech onset from speaker $q$ beginning at time $t_{on} \in \mathbb{R}$. From (\ref{eq:reverbIR}) it arrives at the microphone $i$ via direct-path at time 
\begin{equation} \label{eq:tqi}
t_{qi} = t_{on} + t_{d_{qi}} . 
\end{equation}
%
{\color{black}We assume that the reflections of preceding signals are negligible in comparison with a distinct onset.} 
Also from the reverberation model in Section \ref{section:reverb} we can see that its {\color{black}diffuse} reflections begin to arrive at $t_{qi} + \tau_{qi} $. 
Thus by expanding the convolution $s^{(\hbar)}_q(t) \ast \mathrm{h}_{qi}(t)$ in (\ref{eq:fbsig}),
we can see that at the vicinity of the distinct onset, $x^{(b)}_{i}(t)$ is composed of its direct-path and reflections, i.e.  
\begin{equation} \label{eq:sq}
x^{(b)}_{i}(t) = x^{(b)}_{d_{i}}(t) + x^{(b)}_{R_{i}}(t) , ~t \geq t_{qi} , 
\end{equation}
where the direct-path component is
\begin{equation} \label{eq:dirb}
x^{(b)}_{d_{i}}(t) \triangleq [ {s}^{(\hbar)}_q(t-t_{d_{qi}})\cdot \mathrm{h}_{qi}(t_{d_{qi}}) ]  \ast g^{(b)}(t) ,~t \geq t_{qi} , 
\end{equation}
and {\color{black}from (\ref{eq:reverbIR})} the reflections are
\begin{equation} \label{eq:reflectb}
\begin{aligned}
x^{(b)}_{R_{i}}(t) \triangleq & \big[  \int _ { \tau_{qi} } ^{\infty}  \!\!\!{s}^{(\hbar)}_q(t - t_{d_{qi}} \!\! - \! \tau ) \cdot \mathrm{h}_{qi}(t_{d_{qi}}+ \tau ) d\tau \big] \ast g^{(b)}(t) \\
= & \mathrm{h}_R(t) \ast x^{(b)}_{d_{i}}(t) ,
~ t \geq t_{qi} +  \tau_{qi}
 ,
\end{aligned}
\end{equation}
%
%
%
where $\mathrm{h}_R(t)$ can be viewed as the impulse response:  
\begin{equation} \label{eq:reflectionIR}
\mathrm{h}_R(t) = 
\begin{cases}
0, ~ t < \tau _{qi} \\
\nu_{qi}(t) \cdot 10^{-3 \frac{ t }{T_{60} }}, ~ t \geq  \tau_{qi} , 
\end{cases}
\end{equation}
which represents a linear time-invariant (LTI) system, connecting an arbitrary direct-path signal and its random reflections, for a distinct onset. 

\subsection{Upper Bound of {\color{black}Reflection Level}}

{\color{black}W}e can see from (\ref{eq:reflectionIR}) that the exact values of reflections are unknown without complete knowledge of $\mathrm{h}_{qi}(t)$, especially the $\nu_{qi}(t)$ term. Thus using the fact that $ \mathbb{E} (|\nu_{qi}(t)|) \equiv 1 $, we can instead find an upper bound of the level of reflections which is independent on $\nu_{qi}(t)$.

{\color{black}
Using (\ref{eq:reflectb}) and from Appendix \ref{appen:upperBound}, the level of reflections is
\begin{equation} \label{eq:reflectUpBound}
\mathbb{E} \big( \lfloor {x}^{(b)}_{R_{i}}(t) \rfloor \big) 
\leq  
\tilde{ \mathrm{h}} _R (t) \ast \lfloor x^{(b)}_{d_i}(t) \rfloor 
\triangleq \tilde{x}^{(b)}_{R_{i}}(t) , 
\end{equation}
where $\lfloor \cdot \rfloor$ is the half-wave rectification commonly used \cite{lyon1983computational, meddis1997unitary, tolonen2000computationally}, i.e. $\lfloor x \rfloor = \frac{1}{2} ( x + |x|), \forall x \in \mathbb{R}$. $\tilde{x}^{(b)}_{R_{i}}(t)$ is an upper bound of the level of reflections.} 
$\tilde{ \mathrm{h}} _R (t)$ is the minimum mean square error (MMSE) approximation of $ | { \mathrm{h}} _R (t) | $.
\begin{equation} \label{eq:energyIR}
\tilde{ \mathrm{h}} _R (t) {\color{black} \triangleq \mathbb{E} \big( | { \mathrm{h}} _R (t) | \big) }
= 
\begin{cases}
0, ~ t< \tau_{qi}  \\
10^{-3 \frac{ t }{T_{60} }}, ~ t \geq \tau_{qi} .
\end{cases}
\end{equation}
%

Moreover, since the envelope of a distinct onset is a rising ramp as assumed in Section \ref{section:speechmodel}, the delayed reflections are comparatively small. Thus for the duration of the distinct onset, we have 
\begin{equation} \label{eq:precedence}
x^{(b)}_{i}(t) \approx x^{(b)}_{d_{i}}(t),
\end{equation}
which aligns with the ``precedence effect''\cite{litovsky1999precedence} that speech onsets in microphone signals are dominated by direct-path components.

Therefore from (\ref{eq:reflectUpBound}) and (\ref{eq:precedence}), we have
\begin{equation} \label{eq:reflectionapprox}
\tilde{x}^{(b)}_{R_{i}}(t) \approx  \tilde{ \mathrm{h}} _R (t) \ast \lfloor  x^{(b)}_{i}(t) \rfloor , ~ t \geq t_{qi} +  \tau_{qi} .  
\end{equation}

Note that the upper bound $\tilde{x}^{(b)}_{R_{i}}(t) $ is independent on the $\nu_{qi}(t)$ term of $\mathrm{h}_{qi}(t)$. 
Thus we use it as a consistent threshold for suppressing random reflections and detecting distinct onsets in microphone signals $x^{(b)}_{i}(t)$.

\subsection{{\color{black}Recursive Averaging for Reflection Level}}
\label{sec:recursiveAvg}

In practice, signals are observed at a sampling rate of $f_s $. 
Using the fact that (\ref{eq:energyIR}) is a low-pass filtering LTI process, we thus propose a recursive averaging process to approximate (\ref{eq:reflectionapprox}):
\begin{equation} \label{eq:recursive0}
\begin{aligned}
\bar{x}^{(b)}_i(k) = & \lambda \cdot \bar{x}^{(b)}_i(k-1) + (1-\lambda) \cdot \lfloor {x}_i^{(b)}(k/f_s) \rfloor ,
\end{aligned}
\end{equation}
where $\lambda \,(0 < \lambda < 1) $ is a forgetting factor, and hereafter $k \in \mathbb{Z} ~( k = \mathrm{round} (t \cdot f_s) ) $ is the discrete time index. 
%

From (\ref{eq:recursive0}), the recursive averages after the onset arrival can also be rewritten as: 
\begin{equation} \label{eq:recursive}
\begin{aligned}
\bar{x}^{(b)}_i(k)  
&  = (1-\lambda) \sum \limits _{\ell = {k}_{qi} } ^{k} {\lambda}^{k - \ell } \cdot \lfloor {x}^{(b)}_i(\ell/f_s) \rfloor \\
& = \mathrm{h}_A(k) \ast \lfloor {x}^{(b)}_i(k/f_s) \rfloor ,~  k \geq {k}_{qi}  \triangleq \mathrm{round}(t_{qi}\cdot f_s) , 
\end{aligned}
\end{equation}
where the impulse response is:
\begin{equation} \label{eq:recIR}
\mathrm{h}_A(k) = 
\begin{cases}
0, ~ k< 0 \\
(1 - \lambda) \cdot \lambda^k, ~ k \geq 0 .
\end{cases} 
\end{equation} 
%

Equating the upper bound (\ref{eq:reflectionapprox}) and the recursive averages (\ref{eq:recursive}), we have 
\begin{equation} \label{eq:reflectEqual}
\bar{x}^{(b)}_i(k) = \tilde{x}^{(b)}_{R_{i}}(k/f_s) , 
\end{equation}
which leads to
\begin{equation}
 \mathrm{h}_A(k) \approx \tilde{\mathrm{h}}_R(k/f_s).
\end{equation}
From (\ref{eq:energyIR}) and (\ref{eq:recIR}), we get:
\begin{equation} \label{eq:IRapprox}
\begin{cases}
(1-\lambda)\cdot \lambda^k \approx 0, ~ 0 \leq k < \Delta k_{qi} \\
(1-\lambda) \cdot \lambda^{k} \approx 10^{-3 \frac{k}{T_{60}\cdot f_s}} ,~ k \geq \Delta k_{qi} , 
\end{cases}
\end{equation}
where $\Delta k_{qi} \triangleq \mathrm{round}(\tau_{qi} \cdot f_s)$. 

Thus for $\tau_{{qi}}$ larger than a few milliseconds (as discussed in Section \ref{section:reverb}), from (\ref{eq:IRapprox}) we have for $k \geq \Delta k_{qi}$,
\begin{equation} \label{eq:lambda}
\begin{aligned}
\lambda \approx & (1-\lambda)^{-\frac{1}{k}} \cdot 10^{-\frac{3}{T_{60}\cdot f_s} }  \\
\approx & 10^{-\frac{3}{T_{60}\cdot f_s} } .
\end{aligned}  
\end{equation}
%
%
When $f_s=48000${\color{black}Hz}, we can get
$\lambda = 0.9999$ for $T_{60}=1s$, and $\lambda = 0.99$ for $T_{60}=15ms$. 
It is obvious that $\lambda$ increases as $T_{60}$ increases (stronger reverberation). 
As discussed in Section \ref{section:reverb}, $T_{60}$ can be obtained via measurement or estimation \cite{schroeder1965response, lollmann2010improved}. 
{\color{black}Fig.~\ref{fig:subbandSignals} gives an illustration of the recursive averaging for a subband signal. As shown next, we aim to detect the speech onsets and discard the speech offsets. 
\begin{figure}[!h]
\centering
\includegraphics[width=0.5\textwidth]{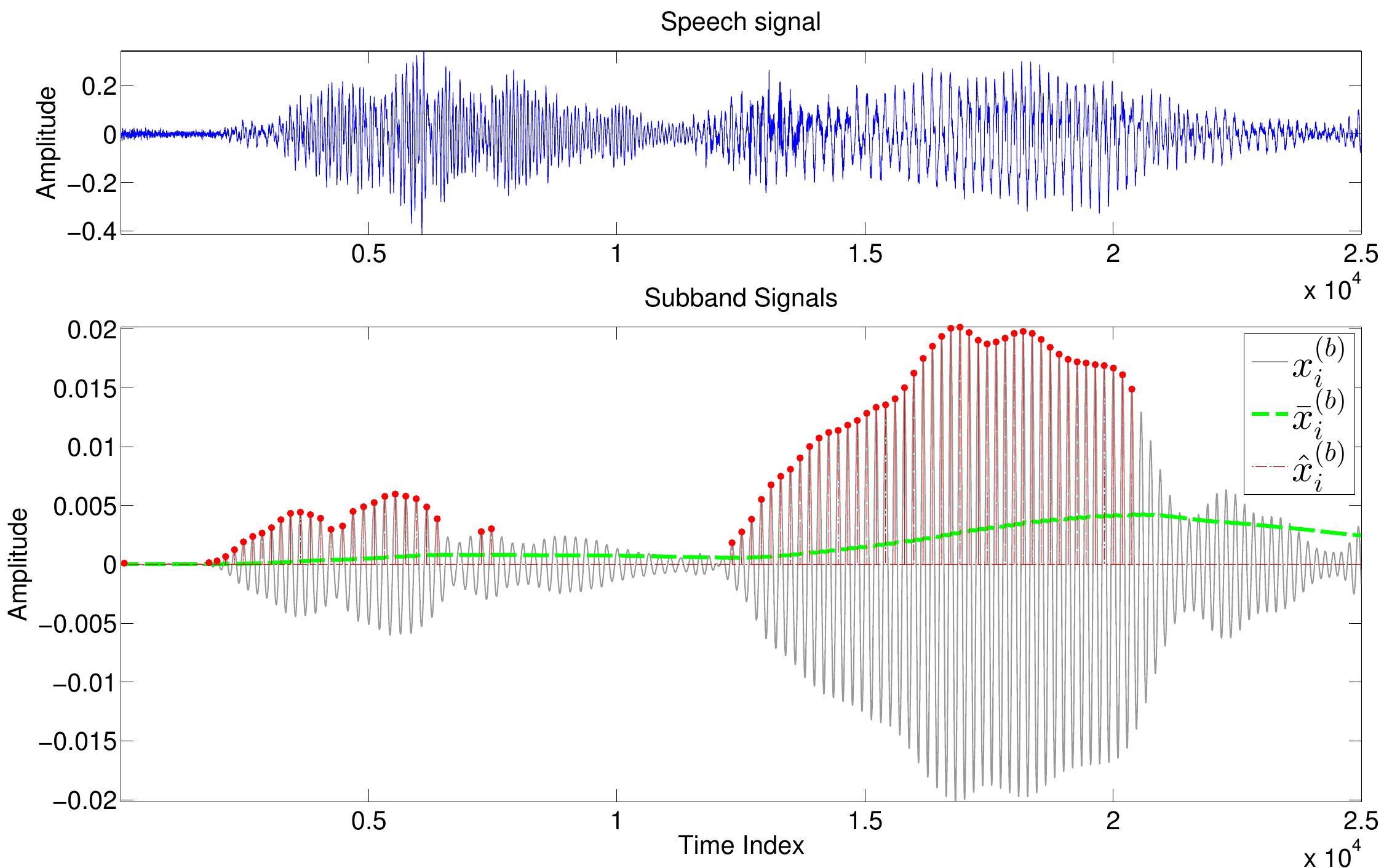}
\caption{\color{black}Speech signal (top panel), subband signal, recursive average, and encoded subband onset signal (bottom panel).}
\label{fig:subbandSignals}
\end{figure}
}
%

%

\subsection{{\color{black}Direct-to-Reflection Ratio and Onset Detection}}
\label{sec:OnsetDetection}

From (\ref{eq:precedence}) and {\color{black}Appendix \ref{appen:dirsubsig}, at the distinct onset we have}
\begin{equation} \label{eq:allSignal}
x^{(b)}_{i}(t) \approx \tilde{S}_{qi}^{(b)}(t)  \cdot \cos ( \tilde{\phi}^{({b})}_{qi}(t)  ) ,~ t \geq t_{{qi}} .
\end{equation}
Thus from (\ref{eq:recursiveConverge}) in Appendix \ref{appen:recursiveAvgPeriod}, we have an upper bound for its recursive averages, i.e.
%
%
\begin{equation} \label{eq:onsetSelect}
\frac{\tilde{S}_{qi}^{(b)}(k/f_s)}{\bar{x}^{(b)}_i(k)}  \geq \pi .
\end{equation}
Note that the {\color{black}equality in} (\ref{eq:onsetSelect}) holds also when the envelope $\tilde{S}_{qi}^{(b)}(k/f_s)$ is constant. Other parts of signals can be speech offsets or weak utterances corrupted by the reflections, hence cannot be used for localization. Thus we discard those parts of signals when (\ref{eq:onsetSelect}) does not hold. The ratio in (\ref{eq:onsetSelect}) can also be viewed as an estimation of the subband direct-to-reflection ratio (DRR). 

The envelope limits the peaks of subband signal.
Thus by comparing the local peaks with the recursive averages according to (\ref{eq:onsetSelect}), we can find the set of indices of distinct onset signals:
\begin{equation} \label{eq:tSNR}
K_{i+}^{(b)} \triangleq \{ {k} ~|~ \cev{k}^{(b)}_{i,n} < {k} < \vec{k}^{(b)}_{i,n} ,~\frac{  \lfloor {x}^{(b)}_i(\hat{k}_{i,n}^{(b)}/f_s) \rfloor  }{\bar{x}^{(b)}_i(\hat{k}_{i,n}^{(b)})} 
\geq {\pi} \} . 
\end{equation}
where index pairs $\cev{k}^{(b)}_{i,n}$ and $\vec{k}^{(b)}_{i,n}$ ($n = 1,2,... $) are zero-crossings that satisfy
\begin{equation}
\label{eq:onsetSelect2}
\lfloor {x}^{(b)}_i(k/f_s) \rfloor > 0 ,\; \forall \; k \in (\cev{k}^{(b)}_{i,n}, \vec{k}^{(b)}_{i,n}) ,
\end{equation}
and $\hat{k}_{i,n}^{(b)}$ is the index of a local peak
\begin{equation} \label{eq:onsetpeaks}
\hat{k}_{i,n}^{(b)} = \argmax\limits_{k}  \lfloor {x}^{(b)}_i(k/f_s) \rfloor  ,~ \forall~ k \in (\cev{k}^{(b)}_{i,n}, \vec{k}^{(b)}_{i,n}) .
\end{equation}

According to (\ref{eq:precedence}), (\ref{eq:reflectEqual}) and (\ref{eq:onsetSelect}), signals with indices $k \in K_{i+}^{(b)}$ are distinct onset signals where direct-path components are dominant, while the rest signals of $ \lfloor {x}^{(b)}_i(k/f_s) \rfloor $ can be corrupted by reflection components and hence are discarded.

{\color{black}\subsection{Onset Encoding}}
\label{sec:encoding}

After we have found the distinct onsets from the subband signals of microphones, we can encode them and then find {\color{black}the} locations of speaker $q$ by estimating the TDOA of direct-path sounds (i.e. differences of $t_{{qi}}$ in (\ref{eq:tqi})) between multiple microphones. 

%
Assuming a slow-changing $\phi^{({\hbar})}_{q}(t)$ in (\ref{eq:speechModel2}), we can rewrite the detected distinct onset signals as a convolution:
\begin{equation} \label{eq:deltaSignals}
\lfloor {x}^{(b)}_i(k/f_s) \rfloor
\approx 
\zeta_{\mathrm{cosine}}^{(\hbar,q)}(k)
\ast 
\sum_{\hat{k}_{i,n}^{(b)} \in \hat{K}_{i+}^{(b)}}  {x}^{(b)}_i(k/f_s) \cdot \delta(k- \hat{k}_{i,n}^{(b)}), 
\end{equation}
where $\zeta_{\mathrm{cosine}}^{(\hbar,q)}(k)$ is the non-negative part of the cosine term with peak at $k=0$,
\begin{equation}
\begin{aligned}
\zeta_{\mathrm{cosine}}^{(\hbar,q)}(k) \triangleq &
\cos \big( \tilde{\phi}^{({b})}_{qi}( ( k + \hat{k}_{i,n}^{(b)} ) /f_s) - \tilde{\phi}^{({b})}_{qi}(\hat{k}_{i,n}^{(b)}/f_s) 
\big), \\ 
& k \in (\cev{k}^{(b)}_{i,n} - \hat{k}_{i,n}^{(b)}, \vec{k}^{(b)}_{i,n} -\hat{k}_{i,n}^{(b)} ) ,
\end{aligned}
\end{equation}
%
the delta function $\delta(k)$ is defined as 
\begin{equation}
\delta(k) =
\begin{cases}
1,~ k = 0 \\
0,~ \mathrm{otherwise}
\end{cases} ,
\end{equation}
and $\hat{K}_{i+}^{(b)} \subset {K}_{i+}^{(b)}$ is the set of indices of onset peaks in (\ref{eq:onsetpeaks}):
\begin{equation}
\hat{K}_{i+}^{(b)} \triangleq \{ \hat{k}_{i,n}^{(b)} ~|~ \frac{  \lfloor {x}^{(b)}_i(\hat{k}_{i,n}^{(b)}/f_s) \rfloor  }{\bar{x}^{(b)}_i(\hat{k}_{i,n}^{(b)})} 
\geq {\pi} \} . 
\end{equation}

Since the precise timing information of onsets (the signal-scaled delta functions in (\ref{eq:deltaSignals})) is crucial to time delay estimation, the slow-changing $\zeta_{\mathrm{cosine}}^{(\hbar,q)}(k)$ term in (\ref{eq:deltaSignals}) can impair the resolution in location estimates (see e.g. \cite{knapp1976generalized}). 
On the other hand however, sharp peaks can be sensitive to errors due to noise and finite observation time \cite{knapp1976generalized}. Thus the choice of encoding the $\zeta_{\mathrm{cosine}}^{(\hbar,q)}(k)$ term for cross-correlation represents a compromise between good resolution, accuracy and reliability.

Considering that localization of multiple concurrent speakers requires good resolution, and assuming that the noise is not too strong, in the scope of this paper, we use the simple way of eliminating the $\zeta_{\mathrm{cosine}}^{(\hbar,q)}(k)$ term and encoding the onsets directly with the scaled delta functions in (\ref{eq:deltaSignals}). 
The resulting signal {\color{black}(cf. Fig.~\ref{fig:subbandSignals})} is denoted as $\hat{x}^{(b)}_i(k)$:
\begin{equation} \label{eq:onsetsignals}
\hat{x}^{(b)}_i(k) = 
\sum_{\hat{k}_{i,n}^{(b)} \in \hat{K}_{i+}^{(b)}}  {x}^{(b)}_i(k/f_s) \cdot \delta(k- \hat{k}_{i,n}^{(b)}), ~ \forall k \in \mathbb{Z}.
\end{equation} 
%
Some other ways of encoding the signals can be found in the literature that generate spikes in-phase with local signal peaks before cross-correlation, however they were inferred from psychoacoustic observations that the neural spikes are generated by the hair cells in the organ of Corti \cite{meddis1986simulation, plinge2010robust}. 
%
%

\section{Multi-speaker Localization}
\label{section:localization}
In Section \ref{section:onset}, we have found the distinct onset signals that are dominated by direct-paths. This section presents our proposed Onset-MCCC localization method using the encoded onset signals. 

\subsection{Onset-MCCC}
\label{section:MCCC}

Assume all speech sources are in far-field.
The TDOA from a source at $\hp{\wp}$ 
to locations of two microphones $\hp{m}_i$, $\hp{m}_j$ can be calculated:
\begin{equation} \label{eq:tauVSp}
\tau_{ji}(\hp{\wp}) = (\| \hp{\wp} - \hp{m}_j\| - \| \hp{\wp} -\hp{m}_i\|) /v ,
\end{equation}
where $\| \cdot \|$ is the Euclidean norm operator, $v=343m/s$ the speed of sound in the environment, and $j = 1,\dots,I$. 
%

We can choose microphone $i=1$ as reference, and align signals from other microphones for each location $\hp{\wp}$:
\begin{equation}
x_{j}(t, \tau_{j1}(\hp{\wp})) = x_j(t + \tau_{j1}(\hp{\wp})) ,
\end{equation}
which can be done in frequency domain:
\begin{equation}
X_{j,\hp{\wp}}(k,f) = X_j(k,f) \cdot e^{-\mathrm{i} 2 \pi f  \cdot \tau_{j1}(\hp{\wp}) } ,
\end{equation}
and then convert the signal back to time domain, as $\tau_{j1}(\hp{\wp})$ may be non-integer. Here $X_{j,\hp{\wp}}(k,f)$ and $X_j(k,f)$ are the short-time Fourier transforms of $x_{j}(t, \tau_{j1}(\hp{\wp}))$ and $x_j(t)$, respectively. $f$ is the frequency, and $\mathrm{i}=\sqrt{-1}$.

Using the proposed onset detection and encoding methods in Section \ref{section:onset}, from (\ref{eq:micSignal}) the resulting subband signals of each aligned microphone signal are denoted as ${x}^{(b)}_j(k, \tau_{j1}(\hp{\wp}))$. Then from (\ref{eq:onsetsignals}), we can obtain the encoded onset signals in each subband, which we denote as $\hat{x}^{(b)}_j(k, \tau_{j1}(\hp{\wp}))$.

Then motivated by the MCCC method \cite{benesty2008microphone}, we can form the aligned signal vector for each filter band:
\begin{equation} \label{eq:sigalVector}
\textbf{x}_b(k,\hp{\wp}) = [ \hat{x}^{(b)}_1(k,0), \hat{x}^{(b)}_2(k,\tau_{21}(\hp{\wp})),..., \hat{x}^{(b)}_I(k,\tau_{I1}(\hp{\wp}))] .
\end{equation}

The spatial correlation matrix \cite{benesty2008microphone} for each subband at location $\hp{\wp}$ can be defined as 
\begin{equation}
\label{eq:spatialCorrMatrix}
\begin{aligned}
R_b(\hp{\wp}) = & \mathbb{E}[\textbf{x}_b^T(k,\hp{\wp}) \cdot \textbf{x}_b(k,\hp{\wp})] 
\end{aligned}
\end{equation}
where $[\cdot]^T$ is the matrix transpose.

We approximate the spatial correlation matrix here {\color{black}using the time average,} by stacking up $\textbf{x}_b(k,\hp{\wp})$ over a short time length of $L \in \mathbb{N}$. 
\begin{equation} \label{eq:stackup}
\boldsymbol{\vec{x}}_b(k,\hp{\wp}) = 
\begin{bmatrix}
\textbf{x}_b(k-L+1,\hp{\wp})- {\bar{\textbf{x}}}_b(k,\hp{\wp}) \\
\vdots \\
\textbf{x}_b(k,\hp{\wp}) - {\bar{\textbf{x}}}_b(k,\hp{\wp})
\end{bmatrix}
\end{equation}
where 
\begin{equation} \label{eq:onlineAvg}
{\bar{\textbf{x}}}_b(k,\hp{\wp}) = \frac{1}{L} \cdot \sum \limits _{k'=k-L+1} ^{k} 
\textbf{x}_b(k',\hp{\wp}) .
\end{equation}
Thus we have {\color{black}the approximation}
\begin{equation}
\label{eq:spatialCorrMatrix3}
\begin{aligned}
R_b(\hp{\wp}) \approx &  \frac{1}{L}\boldsymbol{\vec{x}}_b^T(k,\hp{\wp}) \cdot \boldsymbol{\vec{x}}_b(k,\hp{\wp}) \\
\triangleq & 
\begin{bmatrix}
\sigma^2_{1} & r_{12} & \dots & r_{1I} \\
r_{21} & \sigma^2_{2} & \dots & r_{2I} \\
\vdots & \vdots & \ddots & \vdots \\
r_{I1} & r_{I2} & \dots & \sigma^2_{I}
\end{bmatrix} ,
\end{aligned}
\end{equation}
where we choose $\sigma_i > 0, i=1,...,I$.

Note that subtraction of ${\bar{\textbf{x}}}_b(k,\hp{\wp})$ in (\ref{eq:stackup}) was not proposed in \cite{benesty2004time}, but it is necessary for signals with non-zero averages,
so that we can define the normalized cross correlation coefficient between the $i$-th and $j$-th aligned microphone signals as
\begin{equation}
\label{eq:crossCorr}
\rho_{ij} \triangleq \frac{r_{ij}}{\sigma_i \sigma_j} .
\end{equation}
Therefore, from (\ref{eq:spatialCorrMatrix3}) and (\ref{eq:crossCorr}) we have
\begin{equation}
\label{eq:spatialCorrMatrix2}
\hat{R}_b(\hp{\wp}) \triangleq
\begin{bmatrix}
1 & \rho_{12} & \dots & \rho_{1I} \\
\rho_{21} & 1 & \dots & \rho_{2I} \\
\vdots & \vdots & \ddots & \vdots \\
\rho_{I1} & \rho_{I2} & \dots & 1
\end{bmatrix}=
\Sigma_b^+ \cdot
R_b(\hp{\wp}) 
\cdot \Sigma_b^+ ,
\end{equation}
where the positive semi-definite diagonal matrix
\begin{equation}
\label{eq:sigma}
\begin{aligned}
\Sigma_b = 
\begin{bmatrix}
\sigma_{1} & 0 & \dots & 0 \\
0 & \sigma_{2} & \dots & 0 \\
\vdots & \vdots & \ddots & \vdots \\
0 & 0 & \dots & \sigma_{I}
\end{bmatrix}, 
\end{aligned}
\end{equation}
and $[\cdot]^+ $ is the Moore-Penrose pseudoinverse of a matrix to handle the case when any $\sigma_i=0$ (e.g. for signals with all-zeros or a constant value). Hence in the matrix inverse, the diagonal elements are reciprocals of non-zero $\sigma_i$, while all the rest elements are zeros. It is easy to check that when $\sigma_i=0$ or $\sigma_j=0$, we have $r_{ij}=0$ (by definition), and also $\rho_{ij}=0$ (from the pseudoinverse). We can also see that $\rho_{ij} \in [-1, 1]$ and a peak of $\rho_{ij}$ corresponds to a speaker location as long as microphones are closely located (with no spatial alias). 

Therefore in each subband we have
\begin{equation}
\label{eq:DOA}
\xi^{(b)} (k,\hp{\wp}) \triangleq \prod \limits_{(i,j) \in P^{(b)} }  \lfloor  \rho_{ij}  \rfloor  ,
\end{equation} 
where microphone pairs are selected to avoid spatial alias \cite{plinge2010robust}:
\begin{equation}
\label{eq:micPairb}
P^{(b)} = \{ (i,j) \; | \|m_i - m_j \| < v/(f_c^{(b)} + f_B^{(b)} ) ;  \; i<j \} ,
\end{equation}
$f_c^{(b)}$ and $f_B^{(b)}$ are respectively the center frequency and bandwidth of subband $b$, and we only select microphone pairs with $i<j$ since $\rho_{ij}\equiv \rho_{ji}$ from (\ref{eq:spatialCorrMatrix}) and (\ref{eq:crossCorr}). We can see from (\ref{eq:DOA}) that when $\theta$ matches the DOA of a speaker, ideally $\rho_{ij}$ would be maxima, hence producing a maximum of $\xi^{(b)} $. 

For {\color{black}compact} microphone arrays {\color{black}with a maximal dimension $d$,} {\color{black}let $r$ denote the distance between the array center and the sound source. When $r \gg d$, the TDOAs between microphones are considered independent of $r$.}  We can assume a fixed distance of $r$ {\color{black}to scan and estimate} the DOA $\theta$ ($0^\circ \leq \theta < 360^\circ$) \cite{plinge2010robust}. Thus on the azimuthal plane we have 
\begin{equation} \label{eq:pVStheta}
\hp{\wp} (\theta) = [r\cdot\cos\theta, r\cdot\sin\theta] .
\end{equation}
Note that for simplicity of presentation, we have dropped the variables $b$, $k$ and $\hp{\wp}(\theta)$ for $\sigma_i$, $r_{ij}$ and $\rho_{ij}$ in (\ref{eq:crossCorr}) to (\ref{eq:DOA}).

Finally we can form a location estimation {\color{black} function} using all subband results:
\begin{equation} \label{eq:overallxi}
{\xi}^{\mathrm{onset-mccc}}(k,\theta) = \frac{1}{N_b} \sum \limits _{b=1}^{N_b}  \xi^{(b)}(k,\hp{\wp}(\theta)) , 
\end{equation}
where we can also save computations by using only particular subbands if speaker harmonic frequencies are known \textit{a priori}. 

In this paper, we consider only DOAs of speakers. The proposed method however, can be easily extended for estimating Cartesian locations of speakers using multiple microphone arrays.

\subsection{DOA Estimation}
\label{section:doa}

Considering the sparsity of peaks due to the non-stationarity of speech signals (e.g. pauses, voiced and unvoiced sounds) and interference of reflections, temporal averaging of length $t_{avg}>0$ and time shift of $t_{shift} \in (0,t_{avg}] $ is used{\color{red}:} 
%
\begin{equation} \label{eq:avg_micarray}
\bar{\xi}(k,\theta) =   \frac{1}{f_s \cdot t_{avg}}  \sum_ {k'=k-f_s \cdot t_{avg}+1} ^{k} {\xi}^{\color{black}\mathrm{onset-mccc}}(k',\theta) .
\end{equation}
%

Peaks of $\bar{\xi}(k,\theta)$ correspond to candidate DOA estimates of active speakers. 
For an unknown number of concurrent speakers, we select distinct local peaks as in (\ref{eq:doa1}) and (\ref{eq:doa2}). 
We define $\hat{\Theta}_k$ as the set of DOA estimates at time $k$ that correspond to local peaks of $\bar{\xi}(k,\cdot)$:
\begin{equation} \label{eq:doa1}
\hat{\Theta}_k = \{ \hat{\theta}_{i_k} ~|~ {i_k} = 1,\dots,{N_k} \} ,
\end{equation}
where $\hat{\theta}_{i_k}$ satisfies $\bar{\xi}(k,\hat{\theta}_{i_k}) \geq T_{\bar{\xi}}$ and
\begin{equation} \label{eq:doa2} \color{black}
\hat{\theta}_{i_k} = 
\argmax _{\theta_{i_k} } \bar{\xi}(k,\theta_{i_k}) ,~ \forall \theta_{i_k} \in [\hat{\theta}_{i_k} - \theta_{r},~ \hat{\theta}_{i_k} + \theta_{r}], 
\end{equation}
%
integer $N_k$ is the number of estimated speakers at time $k$. $T_{\bar{\xi}} \in \mathbb{R}$ is an empirical threshold that can be calibrated as the valid range of $\bar{\xi}$ depends on the geometry of the microphone array, the microphones used as well as the noise and interferences. 
Parameter $\theta_r$ ($0 < \theta_{r} \leq 180^{\circ}$) indicates the resolution in estimating DOAs (the minimum angular difference of speakers). A $\theta_{r}$ {\color{black}that is} too small can result in clutter of location estimates, while a $\theta_{r}$ {\color{black}that is} too large can cause miss-detection of DOAs. 
All angles are wrapped into $[0^{\circ}, 360^{\circ})$.  When there is no valid peak found from (\ref{eq:doa2}), we have $N_k = 0$ and $\hat{\Theta}_k = \emptyset$.

{\color{black}
\section{STFT Domain Discussions and the MCC-PHAT} \label{sec:STFT}}

{\color{black}This section discusses the signal models and onset detection algorithms in the STFT domain, as well as our second localization algorithm (MCC-PHAT).

\subsection{Signal Models in the STFT Domain }

Assuming that all signal sources are independent and stationary over short intervals of time, and that the system is linear and time invariant, the signal observed by microphone $i$ can be expressed as
\begin{equation} \label{eq:stftSigMix}
X_i(k,f) = \sum_{q=1}^Q S_q(k,f) \cdot H_{qi}(f) + N_i(k,f) , 
\end{equation}
where $S_q(k,f)$ and $N_i(k,f)$ are STFTs of $s_q(t)$ and $n_i(t)$, respectively. $H_{qi}(f)$ is the Fourier transform of $h_{qi}(t)$.

Applying the TF sparsity assumption \cite{yilmaz2004blind} that for each TF point, at most one source (e.g. $q$) is active, (\ref{eq:stftSigMix}) simplifies to
\begin{equation} \label{eq:stftSigQNoise}
X_i(k,f) \approx S_q(k,f) \cdot H_{qi}(f) + N_i(k,f) . 
\end{equation}
Moreover, when the SNR is high at the specific TF point, (\ref{eq:stftSigQNoise}) further simplifies to (SNR estimation methods are outside of the scope of this paper) 
\begin{equation} \label{eq:stftSig}
X_i(k,f) \approx S_q(k,f) \cdot H_{qi}(f) . 
\end{equation}
Note that similar to the subband decomposition approach (cf. (\ref{eq:micSig}), (\ref{eq:fbsig}) and (\ref{eq:fbsig1})) in Section \ref{sec:subbandDecomp}, the assumptions for (\ref{eq:stftSigMix}), (\ref{eq:stftSigQNoise}) and (\ref{eq:stftSig}) may also become unrealistic in presence of strong reflections. Thus speech onsets are often used for reliable localization. 
Moreover, STFT uses the linear frequency scale, and does not exploit the harmonic structure of speech signals.

\subsection{Onset Detection in the STFT Domain}

Besides other onset detection methods, e.g. spectral flux, phase deviation, etc. as listed in \cite{dixon2006onset} and those in \cite{kuhne2009robust, tho2014robust}, the recursive averaging method in Section \ref{sec:recursiveAvg} and \ref{sec:OnsetDetection} can also be applied to estimate the magnitude level of reflections within each frequency band in the STFT domain, e.g.
\begin{equation}
\bar{X_i}(k,f) = \lambda \cdot \bar{X_i}(k-1,f) + (1-\lambda) \cdot  | X_i(k,f) |  . 
\end{equation}
The onsets can then be found from those TF points with the magnitude higher than the reflection level $\bar{X_i}(k,f)$. 
However, to obtain location estimates from multiple microphones using onset TF points, the clustering techniques \cite{kuhne2009robust, tho2014robust} and the eigendecomposition methods \cite{tho2014robust} often require \textit{a priori} knowledge of the number of sources and may take relatively long observation time to be accurate. Such localization methods may work well for static sources, but are not straightforward for moving speakers.
Moreover, note that there are considerably more frequency bands in the STFT domain, and the speech signals are actually quasi-periodic. These may also negate the benefit of using onset TF points in the STFT domain. 
Thus in what follows we present a reverberation-robust multi-channel localization method in the STFT domain based on the GCC-PHAT method \cite{knapp1976generalized}.}


\subsection{MCC-PHAT}

{The classical GCC-PHAT method {\color{black}uses the cross-power spectrum phase \cite{knapp1976generalized},} for estimating TDOA between two microphones} 
\begin{equation} \label{eq:tdePhat}
\hat{\tau}_{ij}(k) = \argmax _ {\tau} ~ \xi_{ij}^{\mathrm{gcc-phat}} (k,\tau) ,
\end{equation}
where
\begin{equation} \label{eq:ximccphat}
\xi_{ij}^{\mathrm{gcc-phat}} (k,\tau) = 
\int _{-\infty} ^{+\infty} \Xi_{ij}^{\mathrm{gcc-phat}}(k,f) \cdot e^{\mathrm{i} 2 \pi f \tau} ~df ,
\end{equation}
\begin{equation} \label{eq:Xi}
\Xi_{ij}^{\mathrm{gcc-phat}}(k,f) = \frac{G_{x_i x_j}(k,f)}{|G_{x_i x_j}(k,f)|} ,
\end{equation}
and
\begin{equation} \label{eq:Gx1x2}
G_{x_i x_j}(k,f) = \mathbb{E} [ X_i(k,f) \cdot X_j^{\star}(k,f) ] .
\end{equation}
Here $\hat{\tau}_{ij}(k)$ is the estimated time delay between the $i$-th and $j$-th microphones, and $[\cdot]^{\star}$ is the complex conjugate operation. 
%

GCC-PHAT is usually considered sensitive to reverberation because of the anechoic model it relies on \cite{benesty2000adaptive}.
{\color{black}However, s}ince the direct-path component is usually stronger than random reflections as in (\ref{eq:reverbIR}), and also motivated by the idea that the redundant information from multiple microphones can suppress the random reflections \cite{benesty2004time}, we thus formulate 
\begin{equation} \label{eq:mcc-phat}
\xi^{\mathrm{mcc-phat}} (k,\theta) \triangleq  \prod \limits_{(i,j) \in P }  \Bigl\lfloor \xi_{ij}^{\mathrm{gcc-phat}} (k,\tau(\theta))  \Bigr\rfloor ,
\end{equation}
where $\tau$ is a function of $\theta$ from (\ref{eq:tauVSp}) and (\ref{eq:pVStheta}), and the set of microphone pairs $P$ is given in (\ref{eq:micPair}), which is important for avoiding spatial alias in (\ref{eq:Xi}).
\begin{equation}
\label{eq:micPair}
P = \{ (i,j) \; | \| \hp{m}_i - \hp{m}_j \| < v/f_{max} ) ;  \; i<j \} ,
\end{equation}
where $f_{max}$ is the maximum signal frequency considered.
%

%
From (\ref{eq:tdePhat}) and similar to (\ref{eq:DOA}), when $\theta$ matches a speaker DOA at time $k$, $\xi_{ij}^{\mathrm{gcc-phat}}$ would be maxima, hence a maximum of $\xi^{\mathrm{mcc-phat}}(k,\theta) $. 
We refer to this extension of the GCC-PHAT method as the MCC-PHAT method. 
We can see that the GCC-PHAT can be viewed as a special case of MCC-PHAT when there are only two microphones closely placed. 
{\color{black}We recently applied the MCC-PHAT in \cite{lin2018joint} for localization of concurrent speakers in an anechoic condition. This paper details the performance evaluations in reverberant conditions and shows that it is reverberation-robust as expected.}

{\color{black}Similar to (\ref{eq:avg_micarray}), the time average of $\xi^{\mathrm{mcc-phat}} (k,\theta)$ is used for DOA estimation.}
In (\ref{eq:DOA}) and (\ref{eq:mcc-phat}), some other forms of combination, e.g. additive or the t-norm (e.g. the Hamacher product \cite{pertila2008measurement}), may also be used to adjust the range of $\xi$. In this paper we show the results using the simple multiplications. 


\section{Numerical Studies}
\label{section:performance}

In this section we compare the performance of the proposed Onset-MCCC, the extended MCC-PHAT, the Neuro-Fuzzy \cite{plinge2010robust}, the TF-CHB \cite{torres2012robust} and the EB-ESPRIT \cite{teutsch2006acoustic} methods, not only in simulated reverberant and noisy conditions, but also in a real office room ($T_{60}\approx 0.65$s). 
In this paper, circular microphone array is studied for its rotational symmetry. A circular microphone array with $I=8$ equidistant omnidirectional elements and diameter of $d=0.1$m is placed horizontally in the test environment, and speaker DOAs on the same azimuthal plane is evaluated.
The microphone signals are sampled {\color{black}synchronously} at a sampling rate of $48000${\color{black}Hz}. 
Note that it is straightforward to apply the proposed methods with other {\color{black}compact array geometries}. 

\subsection{Experimental Set-up}

We choose $r=1$m in (\ref{eq:pVStheta}) and scan the DOA in $1^{\circ}$ steps. The angular resolution $\theta_r $ in (\ref{eq:doa2}) is chosen as $20^ {\circ}$. 
The frame length in (\ref{eq:stackup}) and (\ref{eq:tdePhat}) is $20$ms. 
For temporal averaging in (\ref{eq:avg_micarray}) the length is $t_{avg}=0.5$s. 
For the MCC-PHAT, the maximum frequency is $f_{max}=3600$Hz. 
In this paper, the TF-CHB and EB-ESPRIT formulate covariance matrices over $20$ms time segments. The TF-CHB and EB-ESPRIT do not require high sampling rate, thus signals are resampled at $8000${\color{black}{Hz}}, and accordingly, they use frequencies up to $4000$Hz. 

For all tests of the Onset-MCCC method in this paper, we choose $\lambda= 0.9998$ in (\ref{eq:recursive0}). We use the gammatone filter \cite{patterson1987efficient,holdsworth1988implementing, CASAwang} as the subband filter in (\ref{eq:micSignal}) for its linear phase and frequency selectivity:
\begin{equation}
\label{eq:gammatone}
g^{(b)}(t) = ({t+t_d})^{\vartheta-1} e^{-2\pi f_b^{(b)} (t+t_d)} \cos(2\pi f_c^{(b)} t ) , ~ t \geq -t_d .
\end{equation}
Here integer $\vartheta$ is the order of filter ($\vartheta=4$ in this paper), $t_d$ is time delay for alignment between filter bands, $  f_b^{(b)}$ scaling factor for the bandwidth \cite{patterson1987efficient, CASAwang}, and $f_c^{(b)}$ the center frequency of each subband chosen on the equivalent rectangular bandwidth-rate scale (ERBS) \cite{CASAwang}.
The center frequencies of the filterbank range from $250$Hz to $3600$Hz, and the number of subbands is $N_b=16$. 

\begin{figure}[!h]
\centering
\includegraphics[width=0.5\textwidth]{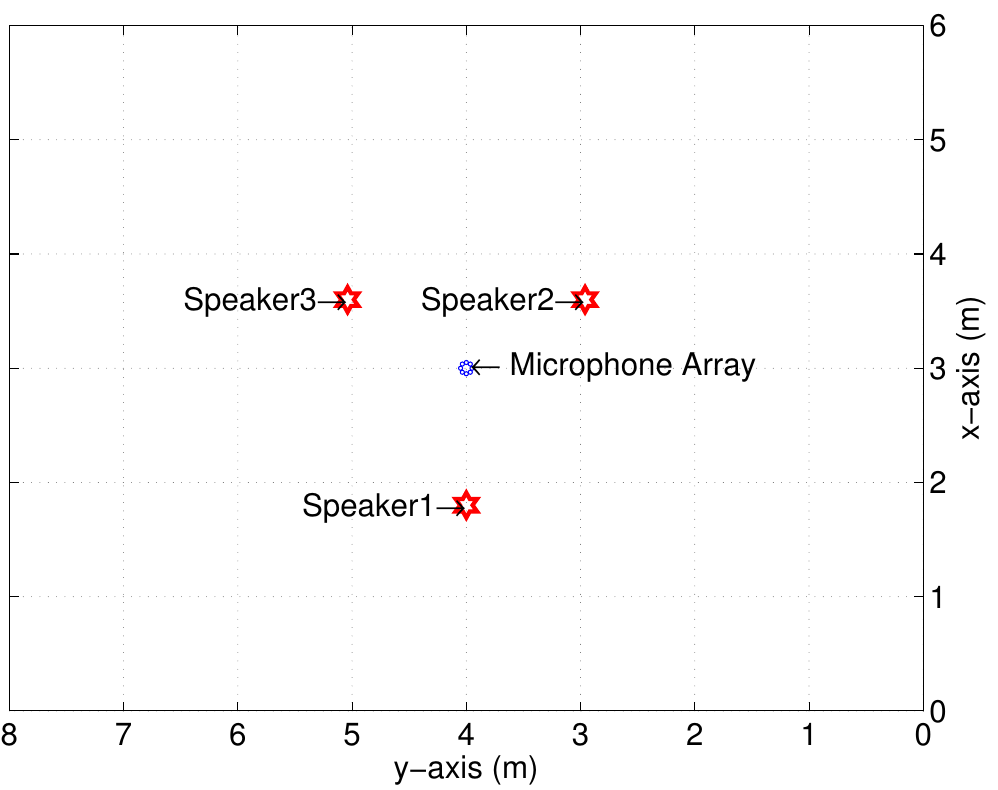}
\caption{Top view of room and set-up (simulation). Locations of microphones and speakers are respectively in  circles and stars.}
\label{fig:roomsetupSimu}
\end{figure}
\begin{figure}[!h]
\centering
\includegraphics[width=0.5\textwidth]{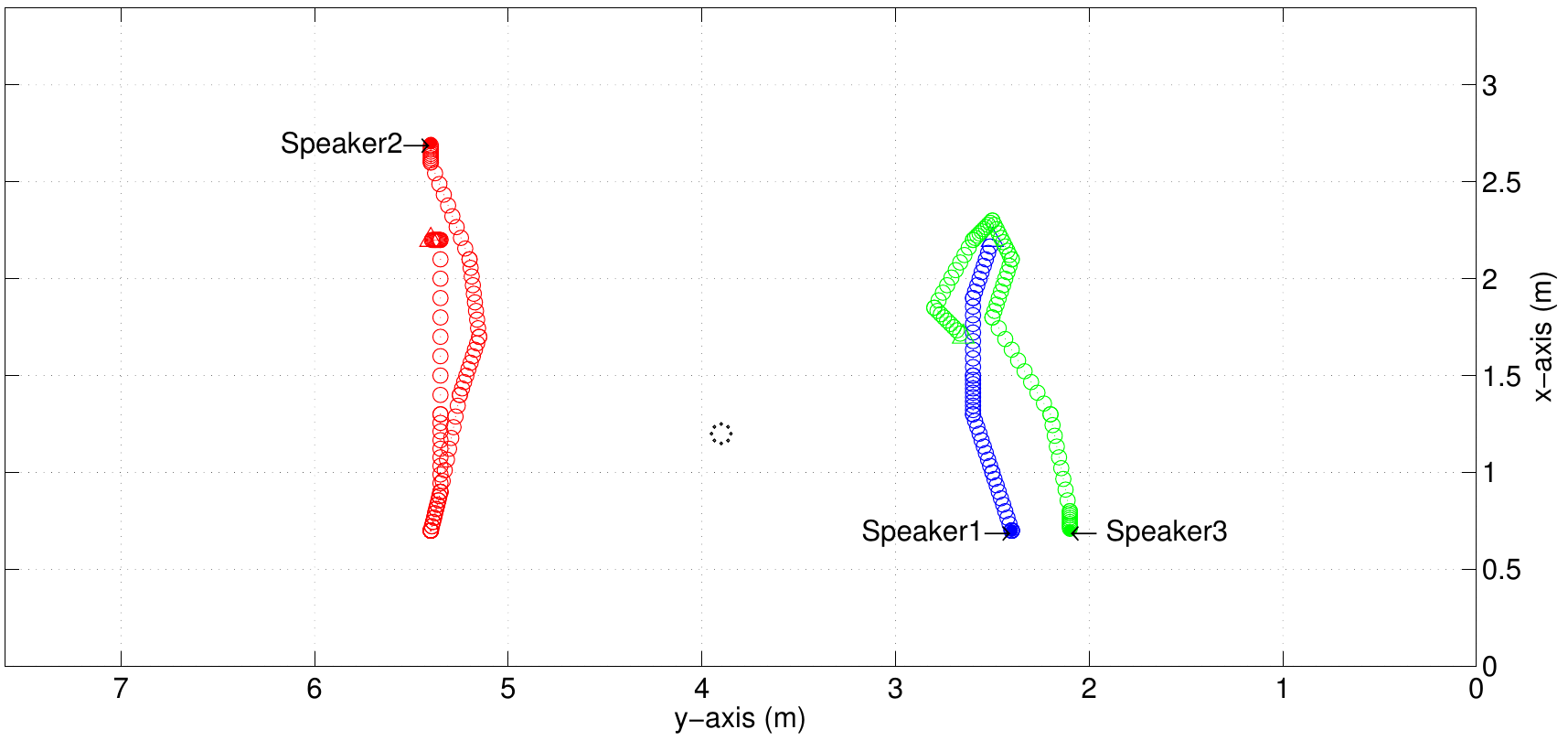}
\caption{Top view of room and set-up (real-world). Locations of microphones are in black circles. Tracks of moving speakers in blue (Speaker1), red (Speaker2) and green (Speaker3). Starting locations of tracks are solid circles and ending locations are triangles.}
\label{fig:roomsetup}
\end{figure}
\begin{figure}[!h]
\centering
\includegraphics[width=0.5\textwidth]{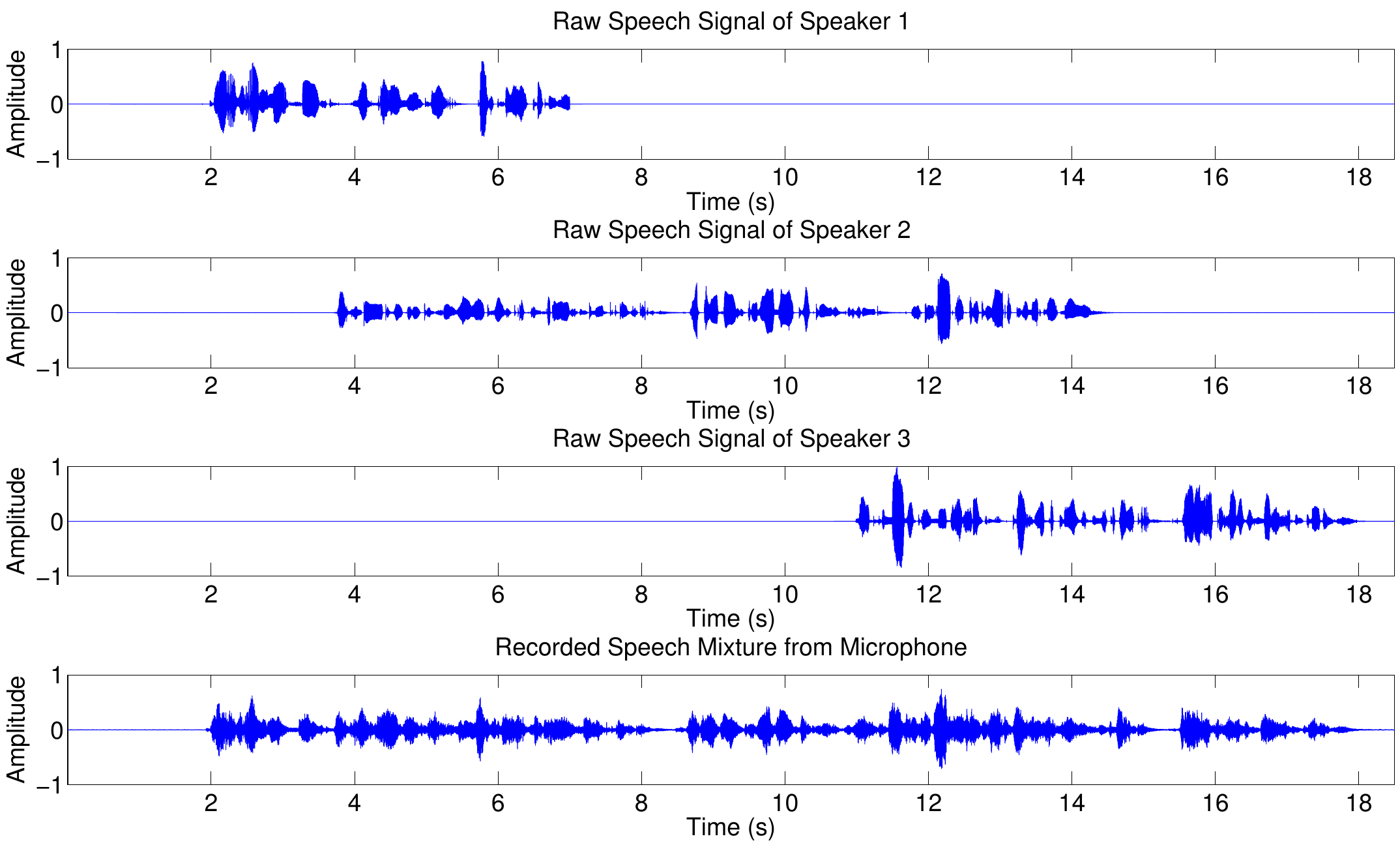}
\caption{Raw signals of moving speakers (top three panels) and a normalized real recording from one of the microphones in the real reverberant room (bottom panel).}
\label{fig:3SpeechSignalsCompare}
\end{figure}
\begin{figure*}[!h]
\centering
\includegraphics[width=\textwidth]{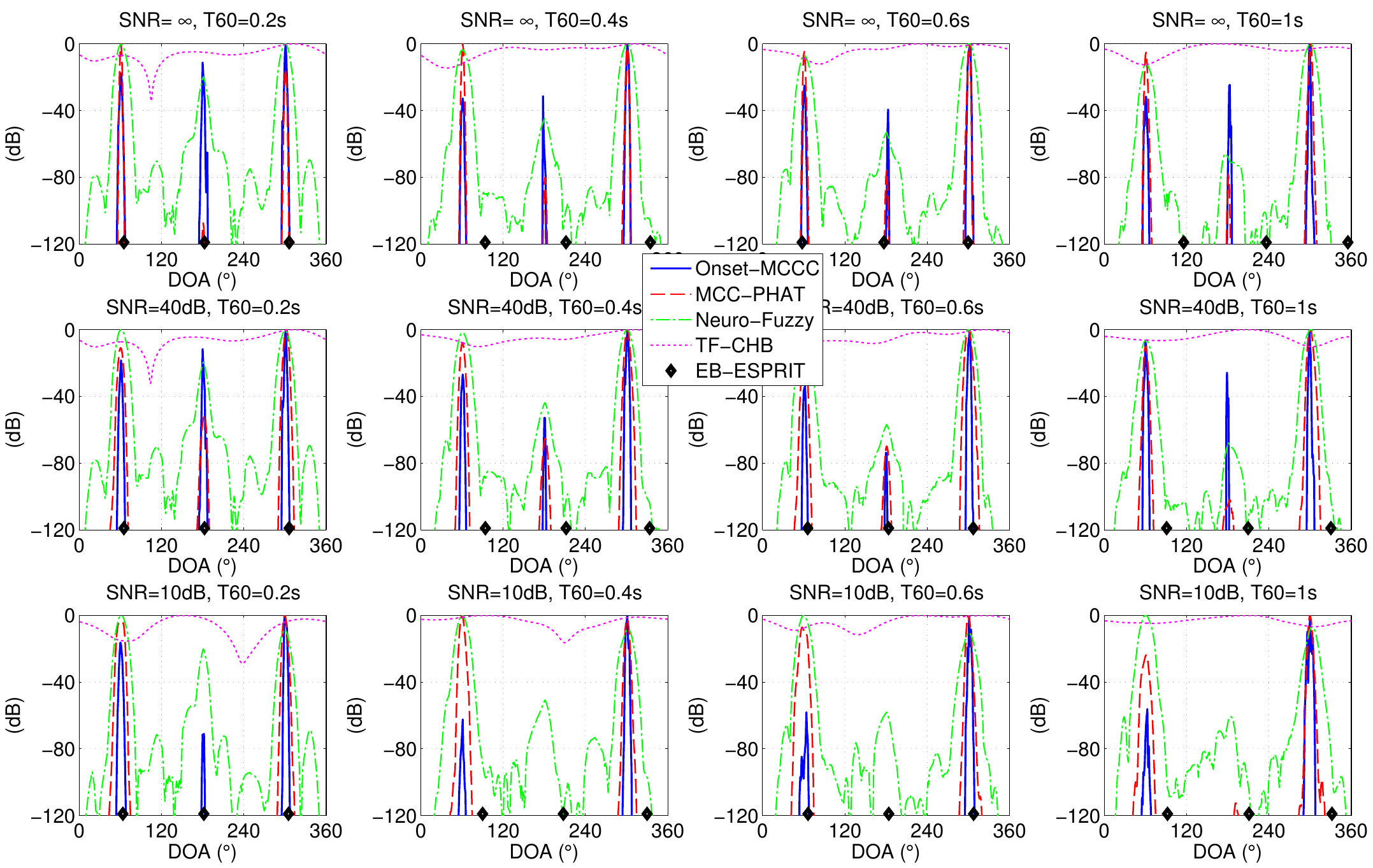}
\caption{Normalized histograms from the Onset-MCCC, MCC-PHAT and Neuro-Fuzzy methods, steered response power from the TF-CHB method and DOA estimates from the EB-ESPRIT method.}
\label{fig:caseAresults}
\end{figure*}

We present the evaluations of {\color{black}four} scenarios in this paper:
\subsubsection{ {Scenario 1} - Three static sources (simulation)} 
As shown in Fig.~{\color{black}\ref{fig:roomsetupSimu}}, three different speakers (Speaker1, male; Speaker2, female; Speaker3, male) are used, each with a speech segment of 4 seconds and the same averaged source power. 
The circular microphone array is placed at the center of a rectangular room of 6m$\times$8m$\times$3m (width$\times$length$\times$height). 
The speakers locate at DOAs of $180^\circ$ (Speaker1),  $300^\circ$ (Speaker2) and $60^\circ$ (Speaker3), respectively, all at a distance of $1.2$m from the center of the microphone array. 
Varying reverberation and additive white noise are applied. The reverberation time $T_{60}$ of the simulated environments ranges from $0.2$ to $1$s simulated using the image-source method (ISM) \cite{lehmann2008prediction}. Additive uncorrelated Gaussian white noise is applied to each microphone and the signal-to-noise ratio (SNR) varies up to 10dB. 

\subsubsection{ {Scenario 2} - Two static sources (simulation)} 
In this case, we study the DOA resolution of respective localization methods. 
The room and microphone set-up is the same as in Scenario 1. Two speakers are located $2$m away from microphone array center at DOAs of $170^{\circ}$ and $190^{\circ}$, respectively. Hence the DOA distance between the two speakers is $20^{\circ}$.

\subsubsection{\color{black}{Scenario 3} - A static source close to wall (simulation)}
{\color{black}In this case, we evaluate the case when a single static source is located close to the wall. The room and microphone set-up is the same as in Scenario 1. 
The DOA is chosen as $45^{\circ}$, and the distance to the closest wall is fixed to $0.2$m. The SNR is $10$dB.
}

\subsubsection{ {Scenario {\color{black}4}} - Three moving sources (real-world)} 
In this case, three speakers are moving while talking. 
The experiment is carried out in a real office room with measured reverberation time of $T_{60}\approx 0.65$s. 
As shown in Fig.~\ref{fig:roomsetup}, the dimensions of the room are 3.4m$\times$7.8m$\times$2.7m (width$\times$length$\times$height). The direction of the x-axis is defined as the $0^{\circ}$ DOA. 
The circular microphone array is placed close to the center of the room at $[1.2, 3.9, 1.5]$m. Omnidirectional electret microphones are used.
Three moving speakers talk and move in a random sequence. 
Speaker signals are chosen from the TIMIT database \cite{garofolo1993timit}. The trajectories of speakers are also plotted in Fig.~\ref{fig:roomsetup}, with different colors. 
Fig.~\ref{fig:3SpeechSignalsCompare} shows the waveforms of speech signals and their starting and ending time, as well as the real recording from one of the microphones in the reverberant room. 

\subsection{Test Results} \label{section:testResults}

\subsubsection{{Scenario 1} - Three static sources (simulation)}
Fig.~\ref{fig:caseAresults} provides the normalized (and scaled by $20 \lg(\cdot)$) DOA estimation histograms of  (\ref{eq:avg_micarray}) from the proposed Onset-MCCC and MCC-PHAT methods as well as that of the Neuro-Fuzzy method, and also the steered-response power of the TF-CHB method and the discrete estimates of the EB-ESPRIT method, over SNR and $T_{60}$. Here we assume that the number of speakers is known \textit{a priori} to the EB-ESPRIT. 

\begin{table*}[t] 
\centering
\caption{RMSE of DOA estimation results {\color{black}(in degrees)} using different methods.}\label{rmseTab}
\begin{tabular}{ @{\extracolsep{5pt}} c c c c c c c c c @{} }
\hline \hline
& \multicolumn{4}{c} {SNR = $\infty$ } & 
	\multicolumn{4}{c} {SNR = $40$dB } \\\cline{2-5}   \cline{6-9}
{Methods} & $T_{60} (s) = 0.2$ & 0.4 & 0.6 & 1 & $T_{60} (s) = 0.2$ & 0.4 & 0.6 & 1 \\
\hline
 Onset-MCCC & 0.82 & 1.41 & 2.52 & 2.16
 & 0.82 & 1.2910 & 1.73 & 0.82 \\  
 MCC-PHAT & 1 & 1 & 1 & 1 & 2.16 & 1 & 1.29 & 1.73  \\ 
 Neuro-Fuzzy &  0.82 & 1 & 1 & 2.45 & 0.82 & 1 & 0.82 & 1 \\
 TF-CHB &  11.09 & 50.60 & 58.61 & -- & 12.33 & -- & 41.64 & -- \\
 EB-ESPRIT & 4.88 & 33.27 & 2.24 & 55.63 & 4.85 & 33.07 & 5.79 & 30.30 \\
\hline \hline 
\end{tabular}
\end{table*}
From Fig.~\ref{fig:caseAresults}, we can see that 
the Onset-MCCC has best resolution (most distinct peaks) overall, and the MCC-PHAT has slightly wider peaks.   
The Neuro-Fuzzy method produces wider peaks compared to both Onset-MCCC and MCC-PHAT. 
The TF-CHB has less distinct peaks in its steered response power.
For the cases of static speakers (Scenarios 1 and 2), the EB-ESPRIT uses the overall average (4 second) of segmental covariance matrices to achieve best accuracy. It has discrete DOA estimates which are plotted in the diamond symbol on the horizontal axes.
Regarding the robustness, the TF-CHB and the EB-ESPRIT methods shows more variances over reverberant conditions. 
TF-CHB produces peaks that correspond to the true speaker locations at low reverberation and low noise (i.e. $T_{60}=0.2$s and SNR of $\infty$ or $40$dB), however, at $T_{60}=0.2$s and SNR=$10$dB, there is a valley close to Speaker3 DOA. The results are worse (peaks do not match speaker DOAs) as the reverberation time further increases. 
The EB-ESPRIT produces reliable DOA estimates at $T_{60}=0.2$s and $T_{60}=0.6$s over SNRs, but has large errors at $T_{60}=0.4$s and $T_{60}=1$s. We can also see that the EB-ESPRIT is robust against white Gaussian noise as this matches the underlying noise subspace model it assumes. Note here that assuming a known number of speakers gives EB-ESPRIT a considerable advantage, as this avoids the errors due to the estimation of the number of speakers. 
In general, the Neuro-Fuzzy shows slightly better robustness at SNR=$10$dB, i.e. it still produces peaks at the location of Speaker1 ($180^{\circ}$) although much weaker than the other two and there are also accompanying spurious peaks. 
At high SNRs ($\infty$ or $40$dB), the proposed Onset-MCCC and the extended MCC-PHAT can reliably detect locations of three speakers, and are also fairly consistent over reverberation levels. 
However, as the noise level increases (SNR=10dB), the peak of Speaker1 (male at $180^{\circ}$) gets much weaker and miss-detected, especially when the reverberation is no less than $0.4$s. This is easy to understand, as the encoding of (\ref{eq:onsetsignals}) leads to superior DOA resolutions, but strong noises can disturb the peaks of subband signals hence the resulting cross-correlation coefficients in DOA estimation. This observation actually follows the discussions on the onset detection and encoding in Section \ref{sec:encoding}. 

The root mean square errors (RMSE) of the DOA estimates using different methods for SNR = $\infty$ and $40$dB are given in Table~\ref{rmseTab}.
We can see that the Onset-MCCC, MCC-PHAT and Neuro-Fuzzy methods achieve similar DOA accuracies within $3^{\circ}$. In general, the accuracy degrades as the reverberation increases. 
The EB-ESPRIT method has overall larger errors comparatively. 
The TF-CHB has largest errors at $T_{60}=0.2$s, and reverberation degrades the performance. Note that TF-CHB does not produce three peaks in some cases (e.g. $T_{60}=1$s), hence do not have valid RMSE results to present. Similarly for SNR=10dB, when there are miss-detections or spurious estimates, the RMSE measure may no longer be consistent or informative, due to the difficulty in mapping the estimates with the ground truth. 
Quantitative accuracy measure for such cases will be given in Scenario {\color{black}4}.

\subsubsection{Scenario 2 - Two Static Speakers (simulation)} \label{sec:case2}

\begin{figure}[!h]
\centering
\includegraphics[width=0.5\textwidth]{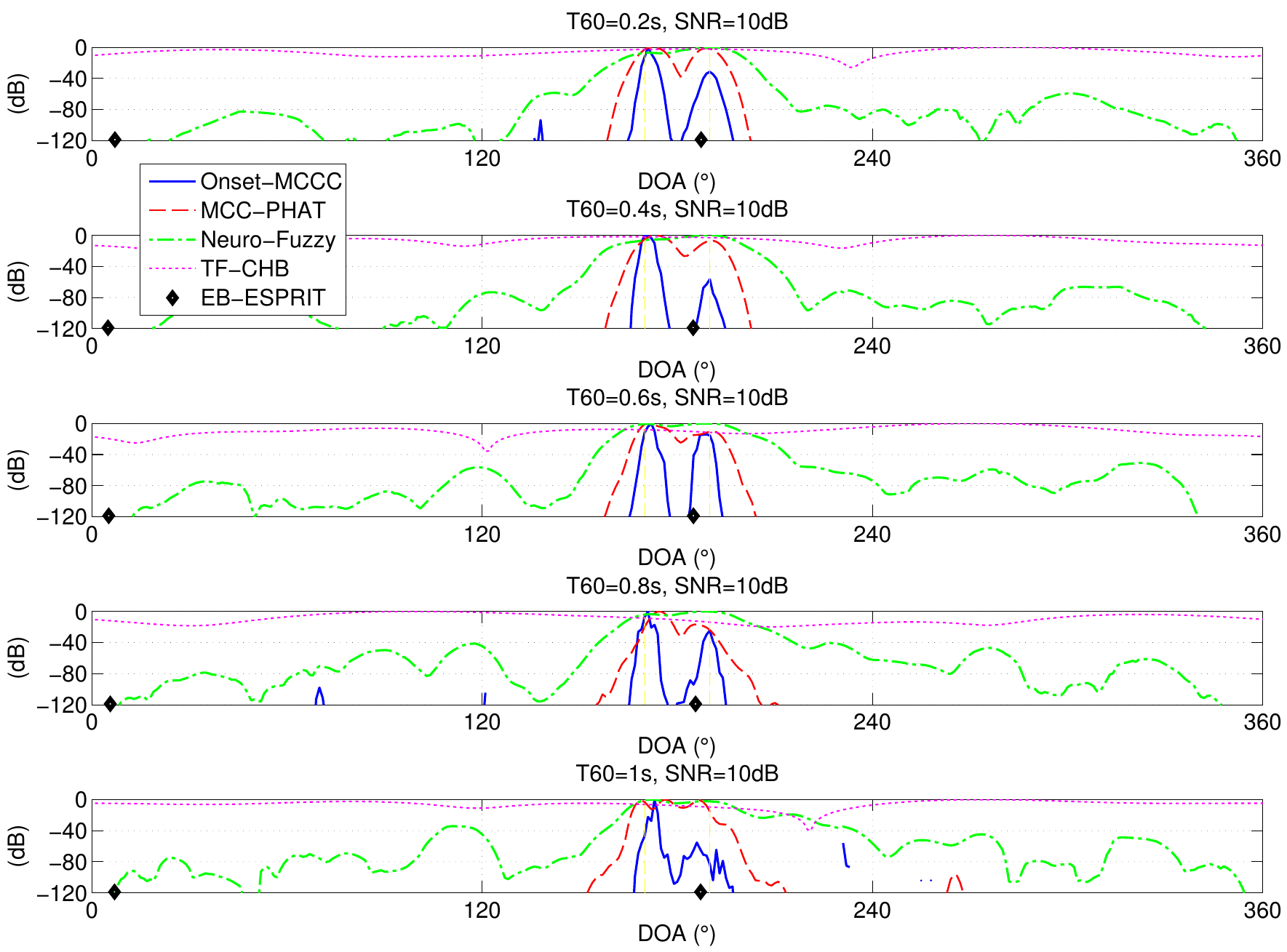}
\caption{Resolution study of different methods at $\mathrm{SNR}=10$dB.}
\label{fig:caseBresults}
\end{figure}

Fig.~{\ref{fig:caseBresults}} plots the DOA localization results for the case when two static speakers locate at DOAs of $170^{\circ}$ and $190^{\circ}$, respectively. The reverberation time ranges from $0.2$ to $1$s. The SNR is $10$dB with additive uncorrelated Gaussian white noise.
It is obvious from the figure that in this case, the Neuro-Fuzzy method no longer form{\color{red}s} two distinct peaks and the two speakers are fused into one speaker in the estimation.
However, the proposed Onset-MCCC and MCC-PHAT methods can reliably form two distinct peaks corresponding to the ground truth. 
{\color{black}Overall}, the TF-CHB forms a wide peak at around $180^{\circ}$, indicating that the two speakers are also fused. 
The EB-ESPRIT again assumes a known number of speakers (which avoids the errors due to estimation of the number of speakers at adverse conditions), and produces a DOA estimate at around $180^{\circ}$ and a second estimate at close to $0^{\circ}$. This indicates that the EB-ESPRIT also has ambiguity to differentiate the two speakers.
In general, as shown in Fig.~{\ref{fig:caseBresults}} {\color{black}(and cf. Fig.~\ref{fig:caseAresults} for a case of larger DOA distances)}, the proposed Onset-MCCC and MCC-PHAT have the best DOA resolution, and the distinct peaks are close to speaker DOAs in most cases. 
Here again, due to the fact that not all methods can always produce the right number of estimates reasonably close to speaker DOAs, RMSE results are not provided. This performance measure problem is addressed next in the Scenario {\color{black}4}. 
%

\subsubsection{\color{black}{Scenario 3} - A static source close to wall (simulation)}
\begin{figure}[!h]
\centering
\includegraphics[width=0.5\textwidth]{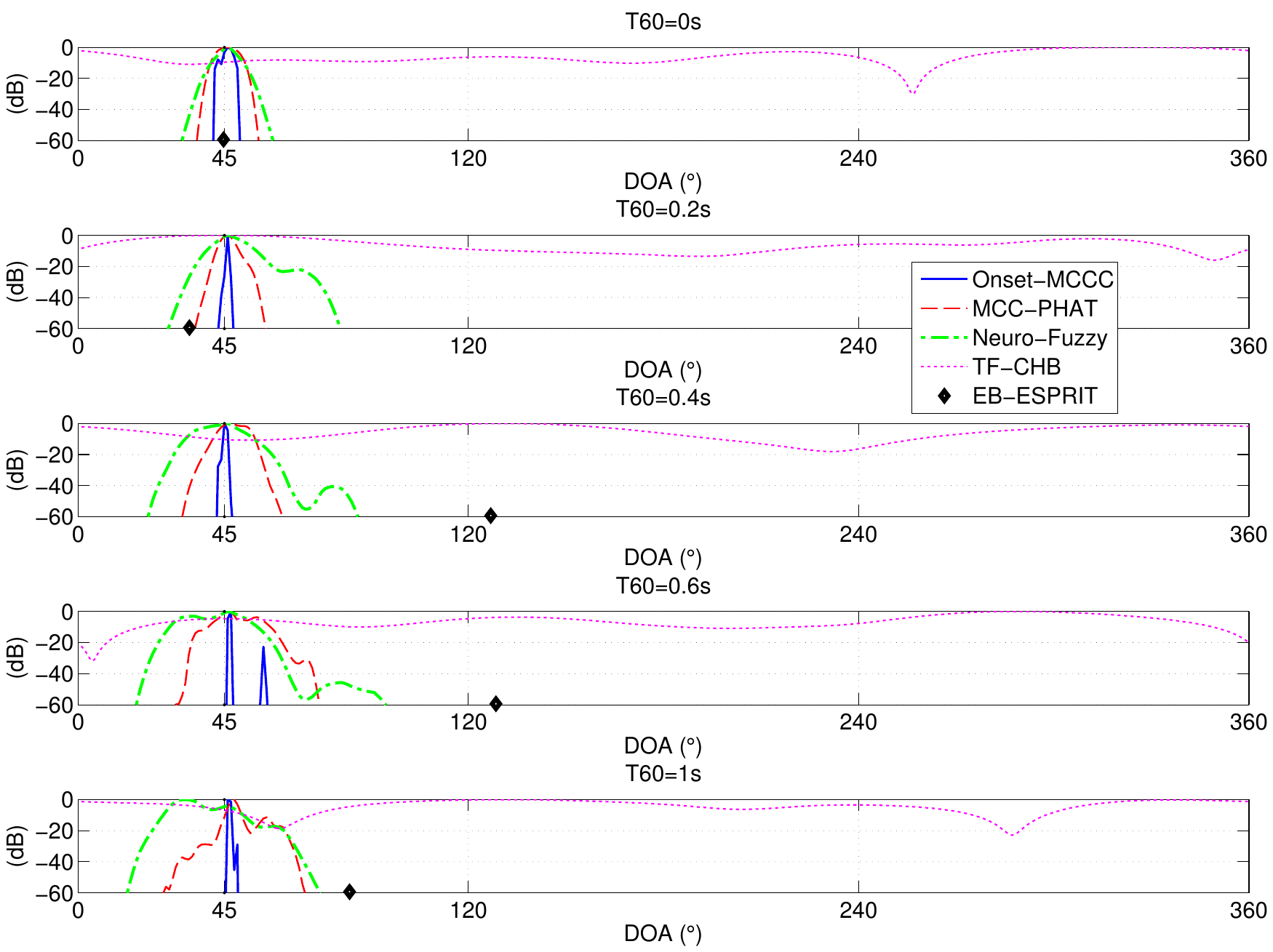}
\caption{\color{black}One static source located at $0.2$m from the wall, at DOA of $45^{\circ}$. SNR=$10$dB.}
\label{fig:DOAresults_1Source_close2wall}
\end{figure}
{\color{black}Fig.~\ref{fig:DOAresults_1Source_close2wall} provides the DOA estimation for a static source located at $0.2$m from the wall. In this case, the early reflection from the closest wall is about $1$ms behind the direct-path. With additive noise at SNR of $10$dB, it is interesting to see that all the methods (except the TF-CHB) work fine at the anechoic case, while as the reverberation increases, the EB-ESPRIT could not find the correct DOAs (given the \textit{a priori} knowledge of one active source).
The Neuro-Fuzzy, Onset-MCCC and MCC-PHAT all produce correct estimates at low to modest reverberation, but have spurious peaks at higher reverberation when the energy of the early reflection is not negligible. 
Note that while the strongest peaks of the Onset-MCCC and MCC-PHAT all correspond to the true speaker DOA, the Neuro-Fuzzy at $T_{60}=1$s does not. Furthermore, the shortest distance from sources to wall in Scenario 2 (cf. Fig.~\ref{fig:caseBresults}) is about 1m, which corresponds to about 6ms time delay between the direct-path and the first early reflection. This follows the assumptions and discussions about the early reflections of the RIR model in Section \ref{section:reverb}. When the source is located close to reflective surfaces (within about 1m), the performance of these localization methods may degrade.}

\subsubsection{Scenario {\color{black}4} - Three Moving Speakers (real-world)} \label{sec:case3}
\begin{figure*}[!h]
\centering
\includegraphics[width=\textwidth]{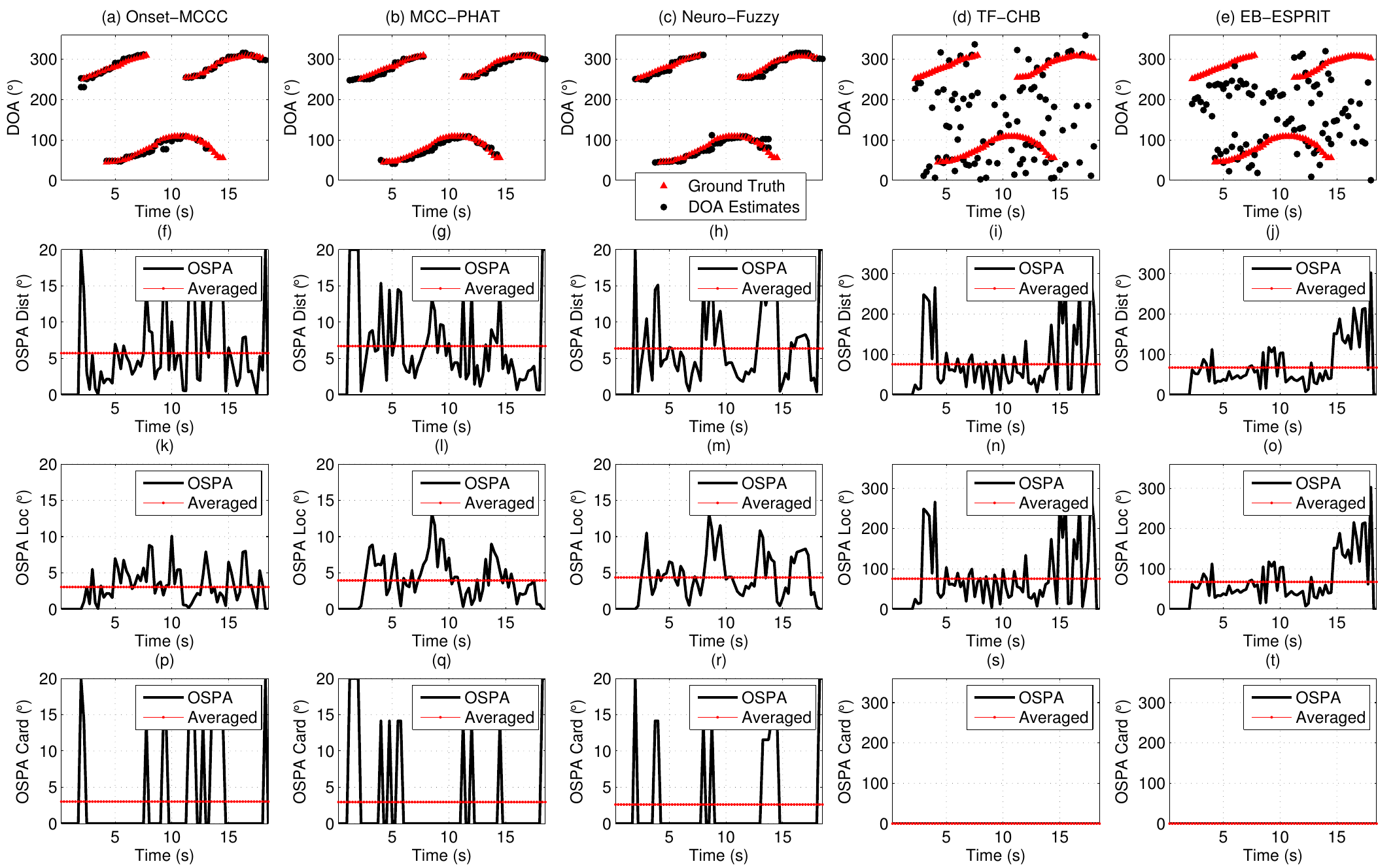}
\caption{DOA estimates (top row) and OSPA results from different methods. Three speakers are moving and speaking in the real reverberant room ($T_{60} \approx 0.65s$).}
\label{fig:DOAresults_MovingSources_2methods}
\end{figure*}
In this case, we estimate the localization of three moving speakers, in a real reverberant room (with measured $T_{60} \approx 0.65$s). 
The trajectories of speakers are given in Fig.~\ref{fig:roomsetup}.
The raw speech signals of the three speakers are given in Fig.~\ref{fig:3SpeechSignalsCompare}. 
%

As noticed in Scenarios 1 and 2, it is neither straightforward nor informative to use the RMSE as the localization performance measure, when there are spurious estimates or miss-detections, especially when there are multiple moving speakers as the locations change over time.  
Therefore, we use the OSPA (optimal sub-pattern assignment) metric \cite{schuhmacher2008consistent} here as a consistent localization performance measure. 
It takes into account the permutations of speakers and evaluates not only the DOA miss-distances (offsets) but also the cardinality errors (errors in the estimated number of speakers). 

The OSPA metric $\bar{d_\rho}^{(c)}$ of two finite sets $\mathit{R}=\{\mathit{r}_1,...,\mathit{r}_m\}$ and $\mathit{S}=\{\mathit{s}_1,,...,\mathit{s}_n\}$, (integers $m \leq n$)  is defined as follows \cite{schuhmacher2008consistent}. (Without causing confusion, we reuse the symbols $n$, $\rho$ and $i$ here.)
\begin{equation} \label{eq:ospa}
\begin{split} 
\bar{d_\rho}^{(c)}(\mathit{R},\mathit{S})  \triangleq  & \Big( \frac{1}{n} \big( \min_{\pi \in \Pi_{n}} \sum_{i=1}^{m} d^{(c)}(\mathit{r}_i, \mathit{s}_{\pi(i)})^\rho + c^\rho(n-m) \big) \Big)^{\frac{1}{\rho}}  ,
\end{split}
\end{equation}
where $\rho \geq 1,~ c > 0$, $ d^{(c)}(\mathit{r},\mathit{s}) \triangleq \min (c, \| \mathit{r} - \mathit{s} \|)$, and $\Pi_{n}$ denotes the set of permutations on $\{1,2,...n\},~ n \in \mathbb{N}$. The distance $\bar{d_\rho}^{(c)}(\mathit{R},\mathit{S})$ stands for a $\rho$-th order per-target error. If $m>n,~ \bar{d_\rho}^{(c)}(\mathit{R},\mathit{S}) = \bar{d_\rho}^{(c)}(\mathit{S},\mathit{R}) $. The order parameter $\rho$ determines the sensitivity to outliers, and the cut-off parameter $c$ assigns weights for cardinality errors (number of miss-detections and spurious estimates). 
Parameter $c$ is chosen to be greater than the maximum estimation deviation of respective localization methods. 
We can see that when choosing $\rho=2$, the OSPA metric can be viewed as an extended RMSE measure that selects closest estimate-truth pairs and includes the ``penalty'' for estimation errors in the number of speakers. Hereafter unless otherwise noted, we choose $\rho=2$ and $c= \theta_r = 20^{\circ}$ so that the cardinality errors are weighted according to the DOA resolution.

The final DOA estimates from (\ref{eq:doa1}) and OSPA results are given in Fig.~\ref{fig:DOAresults_MovingSources_2methods}. 
Each column shows the results from a particular method, i.e. from the left column to the right column, the methods used are respectively the Onset-MCCC, MCC-PHAT, Neuro-Fuzzy, TF-CHB and EB-ESPRIT. 
In the top panel of each column, DOA estimates over time are plotted in black dots, while the ground truth locations of speakers are plotted in red triangles. 
We can see from panels (a) to (e) that although there are several spurious peaks and gaps using the Onset-MCCC, MCC-PHAT and Neuro-Fuzzy methods, they provide close and clean DOA estimates in general. However, the TF-CHB and EB-ESPRIT methods do not produce reliable results in this case of moving speakers. Here the TF-CHB uses short-time histograms over $0.25$s periods and the EB-ESPRIT uses time averages of covariance matrix per $0.25$s. Directly using $20$ms segmental covariance matrices also produces large errors. 
Note that the TF-CHB and EB-ESPRIT methods produce large DOA errors, even though they assume the prior knowledge of the number of speakers over time here. Their estimation results can be worse otherwise. The parameter $c$ is chosen to be $360^{\circ}$ for TF-CHB and the EB-ESPRIT to show the complete OSPA results. It is obvious that these two methods cannot accurately follow the locations of moving speakers in the given scenario. Besides the difficulties in dealing with reverberation, it can also be ascribed to the fact that the EB-ESPRIT method relies on the mathematical expectation of covariance matrix, which can be better approximated using time averages for static speakers but not as much for moving speakers.  

The OSPA results provide closer details of the DOA estimation errors. 
Panels (f) to (j) provide the overall OSPA distances, which are composed of the respective OSPA location errors (panels (k) to (o)) and OSPA cardinality errors (panels (p) to (t)), as shown in (\ref{eq:ospa}). 
Here the OSPA location errors measure the deviations from the location estimates to the ground truth locations. 
We can see from (k) that the proposed Onset-MCCC method has lowest averaged OSPA location error of about $3^{\circ}$, and the maximum error over time is about $10^{\circ}$. The MCC-PHAT and the Neuro-Fuzzy methods as in (l) and (m) produce slightly higher averaged OSPA location errors of about $4^{\circ}$, and the maximum errors over time exceed $12^{\circ}$. From panels (n) and (o), the TF-CHB and EB-ESPRIT methods produce averaged OSPA location errors of about $70^{\circ}$ and $60^{\circ}$, respectively. 
The OSPA cardinality errors measure the cardinality errors in DOA estimates. For example, in panel (p) at time of about $2$s, a spurious estimate when there is no speaker gives an OSPA cardinality error of $c=20^{\circ}$, while in panel (q) at about $5$s, a miss-detection (only one speaker is detected) when there are two speakers gives an OSPA cardinality error of $14.1^{\circ}$. 
The averaged OSPA cardinality errors for the Onset-MCCC, MCC-PHAT and Neuro-Fuzzy methods are all about $3^{\circ}$. 
In panels (s) and (t), since the TF-CHB and EB-ESPRIT assume \textit{a priori} knowledge of the number of speakers over time, they have zero cardinality errors here. 
To sum up, the overall OSPA distances as shown in (f) to (j) demonstrate that the proposed Onset-MCCC and MCC-PHAT methods and the Neuro-Fuzzy can locate moving speakers with an averaged OSPA error of about $6^{\circ}$, while the TF-CHB and EB-ESPRIT methods produce significant localization errors in this case. 

\section{Conclusions} \label{section:conclusion}

This paper proposes a novel reverberation-robust speaker localization method, which we refer to as the Onset-MCCC. 
The Onset-MCCC method first decomposes speech mixtures in TF domain so that harmonic components of speakers do not interfere with each other, based on the speech signal model and the TF sparsity assumption. 
Then it derives a novel onset detection and encoding approach which can detect the direct-paths components from reverberant microphone recordings, based on speech signal and acoustic RIR models. 
Furthermore, it formulates the MCCC of the direct-paths signals while avoiding spatial alias, and produces overall DOA estimates. 
We also implement the MCC-PHAT method, which utilizes the redundant information from multiple closely placed microphones to suppress the impact of reflections. 

Performance of the presented methods is studied using not only simulated signals of reverberation time from $0.2$ to $1s$, but also real recordings in an office room of $T_{60}\approx 0.65s$. Evaluation results show that both the proposed Onset-MCCC and MCC-PHAT localization methods can reliably locate not only static speakers but also multiple competing and moving speakers, in presence of strong reverberation. 
The proposed Onset-MCCC and MCC-PHAT speaker localization methods can also provide {\color{black}good} DOA resolution in presence of reverberation and uncorrelated white noise.
Comparison with other baseline localization techniques {\color{black}in} various reverberant conditions demonstrates the benefits of the proposed methods.

\section*{Acknowledgment}
The author would like to acknowledge the contributions of the Australian Postgraduate Award and Australian Government Research Training Program Scholarship in supporting this research.
Due thanks are given to Prof. Sven Nordholm, Dr. Tze-Chuen Toh, and anonymous reviewers for the constructive review comments on earlier revisions of the manuscript. 

\appendices

\section{Expression of the Direct-path Subband Signal}
\label{appen:dirsubsig}

From (\ref{eq:speechModel2}), the speech harmonic component is{\color{red}\footnote{Considering frequency domain meanings of convolution and complex exponentials for the Fourier transform and inverse Fourier transform.}}
\begin{equation}\label{eq:speechModelenv} 
\begin{aligned}
s^{(\hbar)}_q(t) 
= & A^{(\hbar)}_{q}(t) \cdot \cos \big( {\hbar} \cdot \omega_q \cdot t + \phi^{({\hbar})}_{q}(t) \big) \\
= & \frac{1}{2} {A}^{(\hbar)}_{q}(t) \cdot [ e^{\mathrm{i} {\hbar} \cdot \omega_q \cdot t + \phi^{({\hbar})}_{q}(t)} + e^{- \mathrm{i} {\hbar} \cdot \omega_q \cdot t - \phi^{({\hbar})}_{q}(t)} ]   
\end{aligned}
\end{equation}

Using linear-phase filters, e.g. the gammatone filter, from (\ref{eq:gammatone}) we have
\begin{equation} \label{eq:gammatoneenv}
\begin{aligned}
g^{(b)}(t) 
& = \tilde{g}^{(b)}(t) \cdot \cos(2\pi f_c^{(b)} t ) \\
& = \frac{1}{2} \cdot \tilde{g}^{(b)}(t) \cdot ( e^{\mathrm{i} 2\pi f_c^{(b)} t} +  e^{- \mathrm{i} 2\pi f_c^{(b)} t} )
\end{aligned}
\end{equation}
where 
\begin{equation}
\tilde{g}^{(b)}(t) = ({t+t_d})^{\vartheta-1} e^{-2\pi f_b^{(b)} (t+t_d)}
\end{equation}

From (\ref{eq:speechModel2}) and (\ref{eq:gammatoneenv}), when $ {\hbar} \cdot \omega_q \approx 2\pi f_c^{(b)} $,  the direct-path is given as follows: 
%
\begin{align} \label{eq:cosinederive}
 & {x}^{(b)}_{d_i}(t)
\approx  [ {s}^{(\hbar)}_q(t-t_{d_{qi}})\cdot \mathrm{h}_{qi}(t_{d_{qi}}) ]  \ast g^{(b)}(t)  \nonumber\\
& = \mathrm{h}_{qi}(t_{d_{qi}}) \cdot  [ \frac{1}{2} \cdot \tilde{g}^{(b)}(t) \cdot ( e^{\mathrm{i} 2\pi f_c^{(b)} t} +  e^{- \mathrm{i} 2\pi f_c^{(b)} t} ) ] \nonumber\\
&~~~  \ast  [  \frac{1}{2} {A}^{(\hbar)}_{q}(t-t_{d_{qi}}) \nonumber\\ &~ \cdot [ e^{\mathrm{i} {\hbar} \cdot \omega_q \cdot (t-t_{d_{qi}}) + \phi^{({\hbar})}_{q}(t-t_{d_{qi}})} + e^{- \mathrm{i} {\hbar} \cdot \omega_q \cdot (t-t_{d_{qi}}) - \phi^{({\hbar})}_{q}(t-t_{d_{qi}})} ]  \nonumber\\
& \approx \frac{1}{4} \mathrm{h}_{qi}(t_{d_{qi}}) \cdot    
[ {A}^{(\hbar)}_{q}(t-t_{d_{qi}}) \ast \tilde{g}^{(b)}(t) ] \nonumber\\ 
& ~ \cdot [ e^{\mathrm{i} {\hbar} \cdot \omega_q \cdot (t-t_{d_{qi}}) + \phi^{({\hbar})}_{q}(t-t_{d_{qi}}) } +   e^{- \mathrm{i} {\hbar} \cdot \omega_q \cdot (t-t_{d_{qi}}) - \phi^{({\hbar})}_{q}(t-t_{d_{qi}}) } ]  
\nonumber\\
& =  \frac{1}{2} \mathrm{h}_{qi}(t_{d_{qi}}) \cdot     [ A^{(\hbar)}_{q}(t-t_{d_{qi}}) \ast \tilde{g}^{(b)}(t)  ] \cdot \nonumber\\
&~~~~~~ \cos ( {\hbar}  \omega_q  (t-t_{d_{qi}})  + {\phi}_{q}^{({\hbar})}(t-t_{d_{qi}})  )  \nonumber\\
& =   \tilde{S}_{qi}^{(b)}(t)  \cdot \cos ( \tilde{\phi}^{({b})}_{qi}(t)  ) ,~ t \geq t_{{qi}} , 
\end{align}
%
%
where
\begin{equation} \label{eq:cosineenv}
\tilde{S}_{qi}^{(b)}(t) \triangleq 
 \frac{1}{2} \cdot \mathrm{h}_{qi}(t_{d_{qi}})  \cdot 
  A^{(\hbar)}_{q}(t-t_{d_{qi}}) \ast \tilde{g}^{(b)}(t)  , 
\end{equation}
and
\begin{equation} \label{eq:phaseCosine}
\begin{aligned}
\tilde{\phi}^{({b})}_{qi}(t) 
& =   {\hbar}  \omega_q  (t-t_{d_{qi}})  + \phi^{({\hbar})}_{q}(t-t_{d_{qi}})  . \\
%
\end{aligned}
\end{equation}

Particularly, in the case that the speech harmonic component is narrow-band, and the center frequency is within the pass-band of the filter, (\ref{eq:cosineenv}) can be approximated as
\begin{equation} \label{eq:subbandApprox}
\tilde{S}_{qi}^{(b)}(t) \approx 
 \frac{1}{2} \cdot \mathrm{h}_{qi}(t_{d_{qi}})  \cdot 
  A^{(\hbar)}_{q}(t-t_{d_{qi}}) \cdot \tilde{G}^{(b)}(0) ,
\end{equation}
where $\tilde{G}^{(b)}(f)$ is the Fourier transform of $\tilde{g}^{(b)}(t)$.

We can thus see from (\ref{eq:cosinederive}) and (\ref{eq:subbandApprox}) that the subband direct-path signal is an amplitude modulated sinusoid with slow-changing phase. 

%
%
%
%

\section{Simplification of the Reflection Upper Bound}
\label{appen:upperBound}

From (\ref{eq:reflectb}) we have
\begin{equation}  \label{eq:reflecExpect}
\begin{aligned}
&  \lfloor {x}^{(b)}_{R_{i}}(t) \rfloor 
=  \lfloor \mathrm{h}_R(t) \ast  x^{(b)}_{d_i}(t) \rfloor  \\
& \leq  \lfloor \mathrm{h}_R(t) \rfloor \!\ast\! \lfloor x^{(b)}_{d_i}(t) \rfloor + \lfloor -\mathrm{h}_R(t) \rfloor \!\ast\! \lfloor -x^{(b)}_{d_i}(t) \rfloor 
, 
\end{aligned} 
\end{equation}
where $\lfloor \cdot \rfloor$ keeps the non-negative part of signals while clipping negative signals to zero, i.e. $\lfloor x \rfloor = \frac{1}{2} ( x + |x|), \forall x \in \mathbb{R}$. In the second line of (\ref{eq:reflecExpect}) the equality holds when $\mathrm{h}_R(\tau) \cdot x^{(b)}_{d_i}(t-\tau) \geq 0, ~\forall \tau \geq \tau_{qi} $. 
%

%
%

Thus from (\ref{eq:cosinederive}) and (\ref{eq:reflecExpect}) we have
\begin{equation} \label{eq:reflectLeq}
\begin{aligned}
&  \lfloor {x}^{(b)}_{R_{i}}(t) \rfloor
\leq   \lfloor \mathrm{h}_R(t) \rfloor \ast \lfloor \tilde{S}_{qi}^{(b)}(t)  \cdot \cos ( \tilde{\phi}^{({b})}_{qi}(t)  ) \rfloor \\
&~~~~~~ + \lfloor -\mathrm{h}_R(t) \rfloor \ast \lfloor - \tilde{S}_{qi}^{(b)}(t)  \cdot \cos ( \tilde{\phi}^{({b})}_{qi}(t)  ) \rfloor  \\
& =  \lfloor \mathrm{h}_R(t) \rfloor \ast [ \tilde{S}_{qi}^{(b)}(t)  \cdot \cos ( \tilde{\phi}^{({b})}_{qi}(t)  ) ]_{t \in T_+^{(b)}} \\
&~~~~~~ + \lfloor -\mathrm{h}_R(t) \rfloor \ast [ - \tilde{S}_{qi}^{(b)}(t)  \cdot \cos ( \tilde{\phi}^{({b})}_{qi}(t)  ) ]_{t \in T_-^{(b)}}   \\
& = \big[ \lfloor \mathrm{h}_R(t) \rfloor + \lfloor - \mathrm{h}_R(t) \rfloor \big] \ast [ \tilde{S}_{qi}^{(b)}(t)  \cdot \cos ( \tilde{\phi}^{({b})}_{qi}(t)  ) ]_{t \in T^{(b)}_+}  \\
& = | \mathrm{h}_R(t) | \ast \lfloor x^{(b)}_{d_i}(t) \rfloor 
, 
\end{aligned}
\end{equation} 
where 
\begin{equation} \label{eq:halfwaveApprox}
\begin{aligned}
& [ - \tilde{S}_{qi}^{(b)}(t)  \cdot \cos ( \tilde{\phi}^{({b})}_{qi}(t)  ) ]_{t \in T_-^{(b)}} \\
\approx & [ \tilde{S}_{qi}^{(b)}(t+ \frac{T_b}{2})  \cdot \cos ( \tilde{\phi}^{({b})}_{qi}(t+\frac{T_b}{2})  ) ]_{t \in T_-^{(b)}} \\ 
= & [\tilde{S}_{qi}^{(b)}(t)  \cdot \cos ( \tilde{\phi}^{({b})}_{qi}(t)  ) ]_{t \in T_+^{(b)}} \\
= & \lfloor x^{(b)}_{d_i}(t) \rfloor  ,
\end{aligned}
\end{equation}
$T_+^{(b)}$ is the set of time for non-negative $\cos ( \tilde{\phi}^{({b})}_{qi}(t)  ) $, while  $T_-^{(b)}$ is the set of time for negative $\cos ( \tilde{\phi}^{({b})}_{qi}(t)  ) $.
$T_b$ is the short-term period of $ \cos ( \tilde{\phi}^{({b})}_{qi}(t)  )$.

\section{Recursive Averages of A Periodic Signal}
\label{appen:recursiveAvgPeriod}

\begin{figure}[H]
\centering
\includegraphics[width=0.5\textwidth]{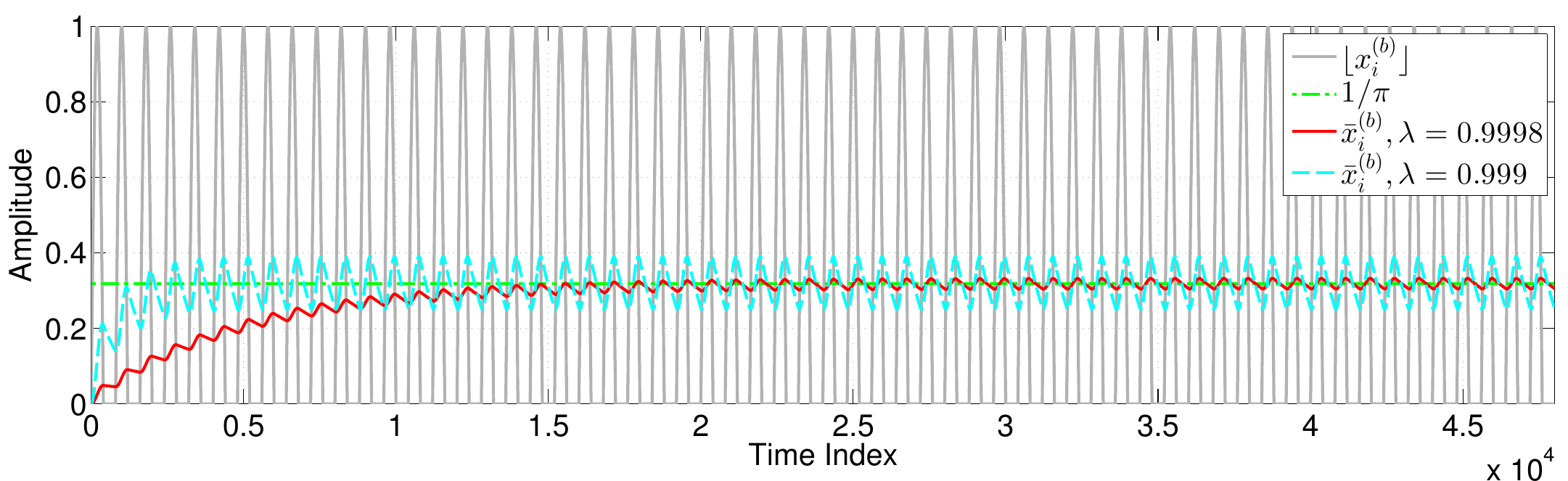}
\caption{\color{black}Intuition for the recursive average upper bound. Amplitude is fixed to 1 for the half-wave rectified sinusoid.}
\label{fig:intuition_recsvAvg}
\end{figure}

We calculate the recursive averages of {\color{black}the half-wave rectified sinusoid} signal, the limit and an upper bound for $\lambda$ less than but close to 1. {\color{black}Fig.~\ref{fig:intuition_recsvAvg} gives an intuition.} 
Assume we have a periodic signal $\lfloor {x}^{(b)}_i(\ell/f_s) \rfloor $ with period of integer $K_b \triangleq \mathrm{round} (T_b \cdot f_s) > 0$, beginning at time $k_0 \in \mathbb{Z}$. From (\ref{eq:recursive}), the limit of the recursive average $\bar{x}^{(b)}_i(k)$ is:
\begin{equation} \label{eq:recsperiod}
\begin{aligned}
& \lim _{k\rightarrow \infty} \bar{x}^{(b)}_i(k) =  \lim _{k\rightarrow \infty} (1-\lambda) \sum \limits _{\ell = k_0 } ^{k} {\lambda}^{k - \ell } \cdot \lfloor {x}^{(b)}_i(\ell/f_s) \rfloor \\
& =  \lim _{k\rightarrow \infty}  \frac{1-\lambda}{K_b}  \sum \limits _{\ell = 0 } ^{k } {\lambda}^{ - \ell\cdot K_b }  \! \sum _{\ell = k_0}^{k_0 + K_b -1 } \!\!\!\!\!\!\! \lambda^{k_0 + K_b -1 - \ell } \lfloor {x}^{(b)}_i(\ell/f_s) \rfloor \\
& = \frac{1-\lambda}{1-\lambda^{K_b}} \cdot \frac{1}{K_b} \sum _{\ell = k_0}^{k_0 + K_b -1 } \lambda^{k_0 + K_b -1 - \ell } \lfloor {x}^{(b)}_i(\ell/f_s) \rfloor .
\end{aligned}
\end{equation}

Using the cosine signal expression as in (\ref{eq:cosinederive}) assuming that $ \tilde{S}_{qi}^{(b)}(t) $ is stable, we can approximate the sum in (\ref{eq:recsperiod}) using the integral, for $\lambda $ close to but less than 1, as in (\ref{eq:lambda}). 
\begin{equation} \label{eq:recsperiod2}
\begin{aligned}
& \int _0 ^{T_b} \!\!\! \lambda^{T_b - t} \cdot \tilde{S}_{qi}^{(b)}(t) \cdot \lfloor \cos ( \tilde{\phi}^{({b})}_{qi}(t) ) \rfloor dt \cdot f_s \\
& \approx  \int _0 ^{T_b} \tilde{S}_{qi}^{(b)}(t) \cdot \lfloor \cos ( \tilde{\phi}^{({b})}_{qi}(t) ) \rfloor dt \cdot f_s \\
& = \tilde{S}_{qi}^{(b)}(t) \cdot \frac{T_b \cdot f_s}{ \pi} .
\end{aligned}
\end{equation}

Thus from (\ref{eq:recsperiod}) and (\ref{eq:recsperiod2}) the limit converges: 
\begin{equation} \label{eq:recursiveConverge}
\begin{aligned}
\lim _{k\rightarrow \infty} \bar{x}^{(b)}_i(k) 
\approx   \frac{1-\lambda}{1-\lambda^{K_b}} \cdot 
\tilde{S}_{qi}^{(b)}(k/f_s) \cdot \frac{1}{ \pi}
\leq  \tilde{S}_{qi}^{(b)}(k/f_s) \frac{1}{ \pi} .
\end{aligned}
\end{equation}
Thus $\tilde{S}_{qi}^{(b)}(k/f_s)/{ \pi} $ is an upper bound of the recursive averages of the signal $\lfloor {x}^{(b)}_i(\ell/f_s) \rfloor $, which is also true when $\tilde{S}_{qi}^{(b)}(k/f_s)$ increases over time (e.g. speech onsets). 



%
%

\ifCLASSOPTIONcaptionsoff
  \newpage
\fi

\bibliographystyle{IEEEtran}
\bibliography{Bibliography}
%

{

%







\end{document}